\documentclass[a4paper,12pt]{article}

\usepackage[english]{babel}
\usepackage[utf8x]{inputenc}
\usepackage[T1]{fontenc}

\usepackage[a4paper,top=2.5cm,bottom=2cm,left=2.5cm,right=2.5cm,marginparwidth=1.75cm]{geometry}

\usepackage{amsmath}
\usepackage{graphicx}
\usepackage{setspace}
\usepackage{apacite}
\usepackage{booktabs}
\usepackage{caption}        
\usepackage{geometry}
\usepackage[titletoc]{appendix}
\usepackage{amssymb}
\usepackage{subcaption}
\usepackage{ragged2e}
\usepackage{tabularx}
\usepackage{bbm}
\usepackage{xcolor}

\usepackage{amssymb}

\usepackage{tikz}
\usetikzlibrary{shapes,decorations,arrows,calc,arrows.meta,fit,positioning}
\tikzset{
	-Latex,auto,node distance =1 cm and 1 cm,semithick,
	state/.style ={ellipse, draw, minimum width = 0.7 cm},
	point/.style = {circle, draw, inner sep=0.04cm,fill,node contents={}},
	bidirected/.style={Latex-Latex,dashed},
	el/.style = {inner sep=2pt, align=left, sloped}
}
\tikzset{
	vertex/.style = {
		circle,
		fill            = black,
		outer sep = 2pt,
		inner sep = 1pt,
	}
}

\tikzstyle{line} = [draw, -latex']
\usetikzlibrary{arrows,decorations.markings,patterns,calc}
\usetikzlibrary{shadows}
\tikzset{shadow scale=1, shadow xshift=-.5ex, shadow yshift=-.5ex,
opacity=.5, fill=black!50, every shadow}

\usepackage{amsmath,amssymb,xcolor}
\usepackage{pgf,tikz}
\usetikzlibrary{arrows,shapes.arrows,shapes.geometric,shapes.multipart,
	decorations.pathmorphing,positioning,shapes.swigs,}

\newcommand\blfootnote[1]{%
  \begingroup
  \renewcommand\thefootnote{}\footnote{#1}%
  \endgroup
}

\newtheorem{ass}{Assumption}[section]

\newtheorem{mydef}{Definition}

\usepackage{booktabs}
\usepackage{threeparttable}
\usepackage{pdflscape}
\usepackage{afterpage}
\usepackage{capt-of}

\usepackage{siunitx}

\title{A Guide to Impact Evaluation under Sample Selection and Missing Data: \\ Teacher's Aides and Adolescent Mental Health
\blfootnote{We would like to thank Dalia Ghanem, Michael Knaus, Elena Mattana, Luke Taylor, and the participants of the Copenhagen University Econometrics Seminar, the IAAE 2023, and the DSSG 2023 for valuable comments and discussion that helped to improve the paper. All remaining errors are ours.}
} 

  \author{ Simon Calmar Andersen \thanks{Aarhus University, Department of Political Science, TrygFonden's Centre for Child Research, Fuglesangs All\'e 4, 8210 Aarhus V, Denmark.} \and Louise Beuchert \thanks{Danish Centre for Social Science Research, Herluf Trolles Gade 11 1052 Copenhagen K, Denmark.} \and Phillip Heiler\thanks{Aarhus University, Department of Economics and Business Economics, TrygFonden's Centre for Child Research, Fuglesangs All\'e 4, 8210 Aarhus V, Denmark, \textit{Corresponding author:} pheiler@econ.au.dk.} \and  Helena Skyt Nielsen \thanks{Aarhus University, Department of Economics and Business Economics, TrygFonden's Centre for Child Research, Fuglesangs All\'e 4, 8210 Aarhus V, Denmark, hnielsen@econ.au.dk} }

\date{This version: \today}

\begin{document}
\renewcommand{\abstractname}{\vspace{-\baselineskip}} 
\maketitle

\doublespacing

\begin{abstract}
\singlespacing
This paper is concerned with identification, estimation, and specification testing in causal evaluation problems when data is selective and/or missing. We leverage recent advances in the literature on graphical methods to provide a unifying framework for guiding empirical practice. The approach integrates and connects to prominent identification and testing strategies in the literature on missing data, causal machine learning, panel data analysis, and more.  
We demonstrate its utility in the context of identification and specification testing in sample selection models and field experiments with attrition. We provide a novel analysis of a large-scale cluster-randomized controlled teacher's aide trial in Danish schools at grade 6. Even with detailed administrative data, the handling of missing data crucially affects broader conclusions about effects on mental health.  Results suggest that teaching assistants provide an effective way of improving internalizing behavior for large parts of the student population.  
\\[4ex]

\end{abstract}
\thispagestyle{empty}
\newpage
\setcounter{page}{1}
\section{Introduction}
Selected samples and missing data are long-standing problems in the evaluation of causal effects and other policy parameters in both observational and experimental studies. Even randomized control trials are often plagued with high levels of non-response or attrition in outcome data. For example, \citeA{ghanem2022testing} find that about 45\% of field experiments in major economic journals have attrition rates higher than the average of 15\%. Some studies even exceed 50\%  and are subject to significant differential attrition rates between treatment and control groups \cite{levy2021social,heiler2022hetbounds}. 
Moreover, instruments for selection are often unavailable. Ignoring selection and missingness can lead to systematically biased estimates and incorrect policy conclusions. 

In this paper, we discuss the use of graphical methods as a unifying framework for causal inference with missing or selectively observed data. In particular, we propose to encode missingness into single world intervention graphs ($m$-SWIGs) as a bridge between graphical methods and counterfactual/potential outcome based approaches.
The main difference to the classic literature on missing data is that missingness and sample selection are themselves understood from a causal or structural perspective and not just as probabilistic events. This allows for a full and transparent encoding of exclusion restrictions and the derivation of implied conditional independencies for both observed variables as well as ``counterfactual'' or ``potential'' variables. The latter can be used for identification, construction of moment functions, and model specification tests. In addition, the graphical toolkit avoids some problems arising from conventional approaches e.g.~when testing for attrition bias and provides a simple tool for deriving sharp testable restrictions implied by nonparametric causal models. We also connect the methodology to more recent developments on canonical frameworks for causal inference in economics and econometrics \cite{heckman2022econometric}. 

This paper provides a general guide for handling evaluation problems with missing data or sample selection without instruments. 
We demonstrate the utility of the graphical approach in the context of point and partial identification in nonparametric sample selection models under varying exclusion restrictions, model specification tests, and testing for attrition in field experiments. The analysis demonstrates that popular proxy tests for no selection or attrition bias tend to be neither necessary nor sufficient for claiming identification. Thus, such tests cannot replace the need for careful comparison of estimands under varying identification assumptions. 
We also make the more general point that, in the presence of an incompletely specified structural model, which is a common practice e.g.~in the literature on panel data methods, derived specification tests tend to either a) exploit too few restrictions, i.e.~have unnecessarily low power, or b) generate non-rejections as a by-product of \textit{accidental independencies}. Identification strategies based on the latter are potentially sensitive to minor changes in the joint distribution and/or measurement, do not easily generalize to conditional effect analysis, and are increasingly hard to justify in setups with complicated interdependencies.  

Based on the single world intervention graphs, we also outline how to construct debiased or orthogonal moment functions that can be used for estimation of both unconditional and heterogeneous effect parameters. Under suitable regularity conditions, they are compatible with the use of machine learning or other non/semiparametric estimation methods for the required nuisance function inputs. In particular, they can allow for the use of random forests, deep neural networks, high-dimensional sparse regression, and other methods for selection and treatment probabilities as well as other quantities. 

Our empirical analysis is concerned with the evaluation of a large-scale field experiment using two types of teacher's aide interventions in grade 6 in the Danish education system. Teacher's aide interventions offer an alternative to more costly and less targeted interventions such as class size reduction. 
The experiment is a cluster-randomised controlled trial across 105 Danish public schools and over 5200 students. Previous analysis suggests that the experiment had a strong positive effect on test scores \cite{andersen2020effect}. In this paper, we are concerned with the average impact on students' mental health. We combine survey data with large administrative data from students and parents 
using unique identifiers. The survey measure for the outcome is subject to significant and differential attrition rates. In particular, relevant attrition rates vary between 10.5\% and 22.0\%. We use the framework suggested in this paper together with novel approaches in the literature on causal machine learning and partial identification to obtain credible estimates of the impact of the intervention on internalizing and externalizing behavior under weak assumptions. 

Our results show that the model used to account for missing data crucially affects conclusions about the effect of teacher's aides on students' internalizing behavior. Under the often-used strong assumption that outcome data are missing completely at random (MCAR) or at random (MAR) (conditional on the observables), effects of teacher's aides are statistically significant and substantial in magnitude (around 7 percent of a standard deviation). Under the weaker assumptions that data are missing not at random (MNAR), estimates are similar but statistically insignificant. Yet, we exploit subgroup heterogeneity in response rates to detect, among other findings, statistically significant and substantial effects of over 12 percent of a standard deviation of teaching assistants on high socioeconomic status students in terms of parental employment even under MNAR. Importantly, in some of these subgroup analyses, we obtain similarly precise and statistically significant effects under the weaker MNAR assumption compared to MAR. 
The empirical study further demonstrates that conclusions from insufficient specification tests do not necessarily align with differences in parameter or interval estimates, in particular when there is heterogeneity across effects and/or selection. 
We discuss the interpretation and implications of these results. 

The paper is structured as follows:
Section \ref{sec_LIT1} discusses the methodological literature. Section \ref{sec_METHOD1} introduces the graphical methods and their relationship to methods in the literature on missing data, sample selection, and causal inference. It also contains applications to sample selection, specification testing, and field experiments with attrition. Section \ref{sec_estimation}  provides examples for debiased estimation. Section \ref{sec_EMPIRICAL1} contains a discussion of the literature with regards to education and mental health as well as the empirical study. Section \ref{sec_CONCL1} concludes.

\section{Methodological Literature} \label{sec_LIT1}
This work is directly related to the literature on missing data and sample selection.
The canonical framework and taxonomy of missing data processes has been pioneered by \citeA{rubin1976inference}, see also \citeA{little2019statistical} for a comprehensive overview. 
Identification of causal effects with or without missing outcomes or sample selectivity is a long-standing problem in economics \cite{heckman1979sample}.
With regards to both, there are two major paradigms in the literature: 1) Model-based approaches that impose assumptions about functional forms such as parametric constraints, additive separability, and/or other distributional restrictions. 2) Design-based approaches that assume knowledge about (treatment) assignment and/or missingness processes and/or their relationship to unobservable variables in the system often framed in relation to a hypothetical (optimal) randomized control trial \cite{rubin2007design}.  
In both cases, identification and estimation is then usually performed via some form of weighting or adjustment-formula \cite{rubin1976inference,pearl1993comment,robins2004effects} and/or explicitly derived specifications only using the observed data distribution, e.g.~control function approaches \cite{heckman1979sample,hausman1979attrition,newey1990semiparametric,matzkin2013nonparametric,wooldridge2015control}.  
Alternatively, for both paradigms, distributions and counterfactual parameters can be bounded under weaker, i.e.~more credible, assumptions using partial identification methods. In particular, \citeA{horowitz2000nonparametric} develop nonparametric bounds for treatment effects in selected samples. \citeA{manski2005partial} consider distributional parameters with missing data.  \citeA{kline2013sensitivity} consider sensitivity analysis and bounds in missing data analysis. There is also a series of papers on bounding subgroup, i.e.~principal strata specific, effects in the sense of \citeA{frangakis2002principal} such as the average treatment effect on the {always-observed}/{always-takers} and more \cite{zhang2003estimation,imai2008sharp,lee2009training,huber2015sharp,semenova2020better,heiler2022hetbounds}. There is also a considerable literature using instrumental variable (IV) based approaches for evaluation and selection, see e.g.~\citeA{imbens1994identification}, \citeA{abadie2003semiparametric}, \citeA{frolich2007nonparametric}, or \citeA{heiler2022efficient}. Our graphical framework can also be applied to an IV context. However, we focus on the case where no classic IVs for selection are available throughout the paper. 

Structural equation models and graphical methods for causal inference have a rich history across multiple disciplines and are becoming increasingly popular in economics, see e.g.~\citeA{pearl2009causaloverview} for a survey and \citeA{heckman2022econometric} for an econometric framework and relationship to alternative approaches. 
The graphical approach to missing data has been pioneered by \citeA{mohan2013graphical} and \citeA{mohan2021graphical}. The key distinction to \citeA{rubin1976inference} is a variable based definition of missingness processes using the logic of structural causal models (SCMs), causal directed acyclical graphs (DAGs), and $do$-calculus. Sample selection using graphical methods is similarly discussed within the SCM/DAG-literature that contains several general but essentially selection on observables identification strategies via $do$-calculus, see \citeA{huendermund2023causal} for a survey. Our discussion regarding identification and specification testing in evaluation problems with missing data is both complementary and partially beyond these approaches and without the need for $do$-calculus.

In the context of panel data models with or without missingness or attrition, identification of parameters has traditionally been based on parametric assumptions \cite{hausman1979attrition,chamberlain1984panel}. Subsequent research often takes a \textit{semi-structural} perspective \cite{manski1987semiparametric,hirano2001combining,altonji2005crossregressors,arkhangelsky2022doubly,ghanem2022testing}. This means that models are assumed to contain some directional and invariant, i.e.~structural, function for the outcome process and then proceed to combine the latter with assumption about the (joint) distribution of unobservables and/or observables such as a treatment variable to achieve identification of effect parameters. With sample selection or missing data on top, a structural relationship is also sometimes imposed for the latter. 
For instance, \citeA{arkhangelsky2022doubly} consider a standard panel setup with a structural function for the outcome and conditional independencies between unobservable confounders and (vector-)treatment assignment.  \citeA{ghanem2022testing} consider a two-period setup with non-separable structural functions for outcome and selection that can depend on both unobservables and observables, in particular a (binary) treatment. They then proceed to impose varying (conditional) independence relationships about the unobservables in both equations and the treatment. 
This semi-structural perspective has also been adopted in generic treatment evaluation and sample selection outside the panel-data context, see e.g.~\citeA{frolich2007nonparametric} or \citeA{huber2013performance}. 
The alternative of imposing a complete structural causal model usually does not affect the identification of main parameters of interest. However, it can yield important different implications for identification of secondary parameters and model validation, e.g.~when testing for attrition bias. We discuss these differences and other problems of the semi-structural approach with a particular focus on treatment evaluation and testing in randomized field experiments under attrition. We show that the graphical approach provides an easy tool to assess the credibility of counterfactual comparisons and to derive more powerful specification tests compared to approaches using statistical independence assumptions. For example, we demonstrate that when testing for internal validity under attrition, our approach can yield more powerful restrictions compared to the semi-structural approach in \citeA{ghanem2022testing} by simply recoding independent treatment assignment in the graph without contradicting the fundamental experimental randomization paradigm.

\citeA{richardson2013single} introduce single world intervention graphs (SWIGs) as a general graphical tool to  bridge the gap between counterfactual/potential outcome logic and graphical models, see also \citeA{shpitser2022multivariate}. Counterfactuals are also well-defined within conventional structural equation models, see e.g.~\citeA{pearl2009causaloverview}. There, however, additional counterfactual independence assumptions are imposed that are, even in principle, experimentally untestable.\footnote{Such assumptions involve probabilistic restrictions of counterfactual variables \textit{across} different SWIGs, so-called \textit{cross-world} assumptions, see Section \ref{sec_exampleMissingness1} for an example using monotonicity.} 
One additional feature is that, in contrast to the DAG/SCM literature, it does not require the use of $do$-calculus to derive adjustment formulas for causal queries and other parameters while retaining much of the useful properties of directed acyclical graphs and related tools such as $d$-separation. In the context of this paper, SWIGs provide a main ingredient for a transparent unifying framework for deriving, evaluating, and testing both observable as well as unobservable dependence relationships in general causal inference problems with missing data.    

The SWIG approach is also closely related to the causal framework considered by \citeA{heckman2015causal} and \citeA{heckman2022econometric} but also applicable outside of the canonical nonparametric structural equation model framework with independent errors. In the context of the latter, the SWIG construction is identical to the hypothetical world model construction considered by \citeA{heckman2022econometric} after conditioning on fixed levels of the intervention variables. 

We combine the missingness graphs with the SWIG approach and show how to use counterfactual adjustment and weighting formula for deriving bias-corrected or orthogonal moment functions for estimation and inference \cite{robinson1988root,robins1994estimation,chernozhukov2018double}. These are important in problems where treatment selection and/or missingness cannot be ignored for identification and the related processes have to be estimated from data.  
Orthogonal moment functions are particularly useful when the researcher is not willing to impose parametric assumption on e.g.~treatment, selection, or outcome process and when there are many confounders. In particular, they can be used with flexible machine learning or other non/semiparametric methods for nuisance function estimation to obtain consistent estimates and valid statistical inference \cite{belloni2014inference,farrell2015robust,chernozhukov2018double,heiler2021effect}, see also \citeA{knaus2022double} for an overview and \citeA{semenova2020better}, \citeA{heiler2022hetbounds}, and \citeA{semenova2023debiased} for extensions to partially identified problems. They are also a main ingredient for flexible heterogeneity analysis in point-identified \cite{lee2017doubly,chernozhukov2018generic,fan2020estimation,semenova2021debiased,heiler2021effect} and partially identified models \cite{heiler2022hetbounds}.

\section{Methodology} \label{sec_METHOD1}
In this section, we formally introduce the graphical models for missing data, the related missingness taxonomy, single world intervention graphs, and their relation to alternative causal frameworks. We provide an overview over some of the basic tools from the literature for deriving independence relationships and to identify counterfactual quantities. We then illustrate the methodology for identification and specification testing in generic nonparametric sample selection models under varying assumptions and in a two-period panel model for experiments with attrition.
\subsection{Graphical Methods}
\subsubsection{Graphical Models with Missing Data}
A missingness graph or \textit{$m$-graph} is a causal directed acyclical graph (DAG) $\mathcal{G} = \mathcal{G}(\mathbb{V},E)$ with nodes $\mathbb{V} = V \cup V^* \cup S \cup U$ and directed edges $E$. $V = V_O \cup V_m$ is the set of observed $V_O$ and (partially) missing nodes $V_m$ with at least one observed record. $V^*$ is a set of proxy variables that relates the selection/non-missingness indicators in $S$ to the underlying (partially) missing variables $V_m$. In particular, for any $v \in V_m$ there are two variables $s_v \in S$ and $v^* \in V^*$, that indicate selection into the data set and are proxy for the missing variable respectively. In particular \begin{align}
	v^* = f(s_v,v) = \begin{cases}
		v &\textit{ if } s_v = 1 \\
		\textit{missing} &\textit{ if } s_v = 0.
	\end{cases} \label{eq_selection-ass1}
\end{align}
Note that we treat any variable in $S$ not as a purely descriptive selection indicator but as a quantity that \textit{causes} or enforces equality between the corresponding variables in $V$ and $V^*$ according to \eqref{eq_selection-ass1}. Thus, $f(\cdot)$ can be considered to be a structural function, i.e.~a directional mapping that is invariant to changes of its inputs. For instance, an outcome in an experiment such as a mental health $v$ could be caused to be selected into the survey according to $s_v$ which itself can depend on other causes such as socio-economic background or other variables in the graph. 
$U$ denotes the set of unobservables. We use bi-directed, dashed edges as a shorthand notation for the existence of unobserved variables that are common parents of two variables in $\mathbb{V}\cup S$. 

Note that our definition is similar to the one suggested by 
\citeA{mohan2013graphical} and \citeA{mohan2021graphical}. We do, however, not require invariance of the distribution of the independent components in $U$ under hypothetical interventions or other manipulations of the graph in Section \ref{sec_SWIGs} and beyond. Moreover, instead of defining a set of missingness nodes, we use non-missingness or \textit{selection} nodes in $S$ to unify notation with the econometric literature on sample selection. 

The resulting $m$-graphs are a very useful tool to represent data-generating processes with structure and missing data and to derive statistical relationships about the involved variables. In particular, they allow for the use of the graphical criterion of $d$-separation to elicit conditional independencies implied by the model: \begin{mydef}[$d$-separation and probabilistic implication  \cite{pearl1988probabilistic}]	~\\ \noindent A path $p$ is  $d$-separated (blocked) by $B$ if \begin{enumerate}
		\item $p$ contains a chain $x \rightarrow b \rightarrow y$ or $x \leftarrow b \leftarrow y$ such that $b \in B$ or 
		\item $p$ contains a collider $x \rightarrow c \leftarrow y$ such that $c \notin B$ and no descendant of $c$ is in $B$.
	\end{enumerate}
	Then $B$ $d$-separates $X$ from $Y$ if it blocks every path from a node in $X$ to a node in $Y$.
	If $X$ and $Y$ are $d$-separated by $B$ in a directed acyclical graph $\mathcal{G}$, then $X$ is independent of $Y$ conditional on $B$, i.e.~$Y\perp X|B$ for every distribution compatible with $\mathcal{G}$.
\end{mydef} 
Thus, given a graph, we can derive necessary conditions that the variables have to obey for the model to be valid.
 Furthermore, given a DAG without further restrictions, $d$-separation is complete, in the sense of that it elicits \textit{all} conditional independencies that can be derived from the model. 
 That means that testing all conditional independencies that follow from $d$-separation jointly is equivalent to testing the \textit{sharp} conditions implied by the model. 


\subsubsection{Single World Intervention Graphs} \label{sec_SWIGs}
A single world intervention graph (SWIG) is a graphical tool to unify causal graphs with potential (or counterfactual) outcome based methodology \cite{richardson2013single}. A SWIG is a modified graph that encodes independencies about factual and counterfactual variables via a simple node-splitting operation. Its main power comes from the fact that $d$-separation retains its completeness property also when applied to SWIGs. Thus, all implied conditional independencies for counterfactual variables implied by the model can be derived from the graph using the same criterion as for standard DAGs. This greatly contributes to the transparency of the source of an identification strategy, unveils potential problems of combining varying exclusion and/or independence relationships, and provides a simple and computationally efficient way to derive conditional independencies for counterfactual variables.

The idea behind the SWIG is a node-splitting operation that 1) introduces a fixed (set of) node(s) $d$ at a level of a hypothetical intervention for variables $D$, 2) moves all edges emanating from $D$ to $d$ instead, and (3) sets the variables of all children of $d$ to their counterfactual values. For a graph $\mathcal{G}$ and intervention $d$, we denote the corresponding SWIG as $\mathcal{G}(d)$.

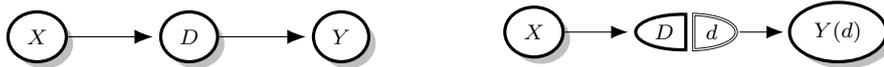
\begin{figure}[!h] \centering \caption{Example DAG and SWIG} \label{fig_DS_example1}
	\begin{subfigure}{0.4\textwidth}
		\begin{tikzpicture}
			\tikzset{line width=1.5pt}
			\node[ellipse,draw,line width = 1.2pt, drop shadow, fill = white] (0) at (-2,0) {\scriptsize$X$};
			\node[ellipse,draw,line width = 1.2pt, drop shadow, fill = white] (1) at (0,0) {\scriptsize$D$};
			\node[ellipse,draw,line width = 1.2pt, drop shadow, fill = white] (2)  at (2,0) {\scriptsize$Y$};
			
			\path (0) edge (1);	
			\path (1) edge (2);	
		\end{tikzpicture}
	\end{subfigure}
	\begin{subfigure}{0.4\textwidth}
		\begin{tikzpicture}
			\tikzset{line width=1.2pt,
				swig vsplit={gap=3pt,
					inner line width right=0.5pt}}
			\node[ellipse,draw,line width = 1.2pt, drop shadow, fill = white] (0) at (-2,0) {\scriptsize$X$};
			\node[name=1,shape=swig vsplit, drop shadow, fill = white] at (0,0) {
				\nodepart{left}{\scriptsize$D$}
				\nodepart{right}{\scriptsize$d$} } ;
			\node[ellipse,draw,line width = 1.2pt, drop shadow, fill = white] (2)  at (2,0) {\scriptsize$Y(d)$};
			
			\path (0) edge (1);
			\path (1) edge (2);	
		\end{tikzpicture}
	\end{subfigure}
\end{figure}

Figure \ref{fig_DS_example1} contains an example of a simple DAG and the corresponding SWIG where we intervene on the variable $D$ by setting it exogeneously to $d$. Such as SWIG reflects a counterfactual scenario under a hypothetical change that leaves the otherwise invariant causal system intact. Thus, these hypotheticals have a genuine causal interpretation. For example, the SWIG $\mathcal{G}(d)$ in Figure \ref{fig_DS_example1} could represent a hypothetical intervention that sets a treatment such as educational resources $D$ to $d$ even though selection into this treatment is based on some characteristics $X$ (socio-economic background, ability etc.) in the observational world. The outcome in the hypothetical world $Y(d)$ is then a counterfactual or potential outcome, e.g.~mental health at intervention input level resources $d$. 

It is important to note that $d$ is treated as a fixed node in $\mathcal{G}(d)$ and not as a random variable, i.e.~any path through $d$ is blocked by construction. This is relevant when characterizing paths and applying $d$-separation to elicit independence relationships. The DAG here implies the factual independence $X\perp Y|D$ while the SWIG implies the counterfactual independence $Y(d) \perp X,D$. Moreover, any SWIG is always defined for its specific intervention. However, since the intervention can take multiple values, the procedure applies to the general \textit{template} of setting $D=d$ for any value $d$ that $D$ is supported on. For example, if $D$ is a binary variable, the single world intervention template in Figure \ref{fig_DS_example1} yields $Y(d) \perp X,D$ for $d=0,1$. There are no assumptions about the dependence of counterfactuals \textit{across} different intervention worlds imposed. For example, the joint distribution of variables $Y(d)$ and $Y(d')$ derived from SWIGs $\mathcal{G}(d)$ and $\mathcal{G}(d')$ as in Figure \ref{fig_DS_example1} are unrestricted contrary to conventional SCMs \cite{richardson2013single}.

We suggest to combine the node-splitting operation with $m$-graphs to obtain what we refer to as missingness single world intervention graphs or $m$-SWIGs. The resulting $m$-SWIGs yield a unifying framework for combining and assessing counterfactual based identification with missingness and/or independence assumptions on observables and unobservables. It also avoids much of the $do$-calculus for identification/recovering causal parameters and replaces it with conventional probabilistic operations on counterfactual variables under the SWIG assumptions.\footnote{\citeA{heckman2022econometric} describe the $do$-calculus as ill-defined. We do not necessarily subscribe to this notion. However, manipulating counterfactual variables under the SWIG and the \citeA{heckman2022econometric} models requires only standard probability theory.}  
For example, simple counterfactual (regression) adjustment and weighting for identification of a potential outcome is given by: \begin{mydef}[Counterfactual adjustment and weighting] \label{def_cfa_ipw1}
	If $Y(d) \perp D | X$ is implied by the SWIG $\mathcal{G}(d)$, then \begin{align*}
		E[Y(d)] = E[E[Y|X=x,D=d]]
	\end{align*} 
    and \begin{align*}
		E[Y(d)] = E\bigg[\frac{Y\mathbbm{1}(D=d)}{P(D=d|X)}\bigg]
	\end{align*}
\end{mydef}
This is equivalent to standard adjustment for confounders and weighting using conditional independence $Y(d) \perp D | X$ in the classic Rubin-Neyman/unconfoundedness framework. The difference is that we start with a transparent graphical or structural model and derive the conditional independence instead of assuming it ex ante. Adjustment and weighting can also be applied to continuous treatments $D$ by using a (smoothed) localized version of the formula above \cite{colangelo2020double}.

\paragraph{Remark (Relationship to hypothetical model approaches):}
The node-splitting operation and general structure of SWIGs bears close resemblance to the extended Haavelmo framework of hypothetical worlds as recasted by \citeA{heckman2015causal} and \citeA{heckman2022econometric}. They focus on a generalized Roy model for exposition, but their ideas extend beyond this framework. In particular, given a nonparametric structural equation model, \citeA{heckman2022econometric} apply an equivalent operation to single node-splitting by adding a modified or \textit{hypothetical} nodes $\tilde{D}$ tied to an original nodes $D$ while simultaneously moving all edges emanating from $D$ to other nodes to the new $\tilde{D}$. They then condition on $\tilde{D} = {d}$ to obtain causal or counterfactual quantities under the hypothetical model. This conditioned, hypothetical world is essentially a SWIG with equivalent intervention. The main difference is that the hypothetical world approach also assume an invariant distribution of exogenous unobserved variables in $U$ as in SCMs. In contrast to the $do$-calculus based approach, both the hypothetical world and SWIG method, does not require other auxiliary constructions for identification and adjustment of quantities that condition on the intervention variables themselves such as the average treatment effect on the treated \cite{shpitser2009effects,richardson2013single,heckman2015causal}. 




\subsection{Selection and Missingness Mechanisms} \label{sec_methodology1_selandmiss}
\subsubsection{Missingness Taxonomy for Variables and Counterfactuals}
We adopt the variable based taxonomy of missingness methods introduced by \citeA{mohan2013graphical}. We use the terms selection/non-missingness interchangeably in this section whenever it does not cause confusion.\\
\textbf{Missing Completely at Random (MCAR)}: The distribution of selection indicators is independent of the (partially) missing variables themselves and any other variable, i.e.~$S \perp (V_O\cup V_m \cup U)$.  \\
\textbf{Missing at Random (MAR)}: The distribution of the selection indicators is independent of the (partially) missing variables themselves and unobservables given the observables, i.e.~$S \perp (V_m \cup U) |V_O$. \\
\textbf{Missing Not at Random (MNAR)}: Neither MCAR nor MAR holds. 

These variable based definitions are slightly stronger than the original, event based definitions by \citeA{rubin1976inference}. However, they allow for a more transparent encoding and are easier to assess in models with multiple or complicated dependencies. 
These basic definitions are with respect to a full graph $\mathcal{G}$. However, one can easily construct cases where e.g.~MCAR/MAR holds for a subset of variables in $S$ with regards to elements in $V_m \cup U$, while for the remaining variables in $S$ the data might fall under the MNAR rubric. We use this subset specific missingness classification throughout this paper when referring to a particular single or selection of variables.

\subsubsection{Missingness of Counterfactuals in SWIGs}
The variable based taxonomy applies equivalently to counterfactuals in a SWIG. However, there are two distinctions compared to a standard DAG. First, missing variables can be counterfactuals. This is best understood as observing (partially) missing variables that would be realized in a hypothetical world where we intervene according to the SWIG. Second, selection indicators themselves can be counterfactual quantities. In particular, if an intervened upon variable causes selection, one can only observe the corresponding potential or counterfactual selection in the hypothetical world.  

\subsubsection{Examples of Missingness Mechanisms} \label{sec_exampleMissingness1}
In the following, we outline three prototypical examples for selection processes that are in the MCAR, MAR, and MNAR category respectively. They are related to classical sample selection models with varying assumptions on exclusion and independence. We provide their causal DAGs, derived SWIGs and implied conditional independencies that can be used for i) testing necessary independence conditions and ii) identification of causal or counterfactual parameters. In addition, we demonstrate that, even when there is no point identification, $m$-SWIGs can still be exploited to derive conditional independencies that are used for set identification methods (often in conjunction with other functional form and/or shape restrictions).

\begin{figure}[!h] \centering \caption{Models $M_1$, $M_2$, and $M_3$: $m$-DAGs and $m$-SWIGs} \label{fig_DS_examplesMiss1}
	\begin{subfigure}{0.4\textwidth}\centering 
		\begin{tikzpicture}
			\tikzset{line width=1.5pt}
			
			\node[ellipse,draw,line width = 1.2pt, drop shadow, fill = white] (1) at (0,0) {\scriptsize$D$};
			\node[ellipse,draw,line width = 1.2pt, drop shadow, fill = white] (2)  at (3.6,-1.2) {\scriptsize$S$};
			\node[ellipse,draw,line width = 1.2pt, drop shadow, fill = lightgray] (3) at (1.2,-1.2) {\scriptsize$Y$};
			\node[ellipse,draw,line width = 1.2pt, drop shadow, fill = white] (4) at (2.4,-2.4) {\scriptsize$Y^*$};
			
			\path (1) edge (3);
			\path (2) edge (4);
			\path (3) edge (4);	
		\end{tikzpicture}\caption{\footnotesize Model $M_1$: MCAR $m$-DAG}
	\end{subfigure}
	\begin{subfigure}{0.4\textwidth}\centering
		\begin{tikzpicture}
			\tikzset{line width=1.2pt,
				swig vsplit={gap=3pt,
					inner line width right=0.5pt}}
			
			\node[name=a1,shape=swig vsplit, drop shadow, fill = white] at (0,0) {
				\nodepart{left}{\scriptsize$D$}
				\nodepart{right}{\scriptsize$d$} } ;
			\node[ellipse,draw,line width = 1.2pt, drop shadow, fill = white] (2)  at (3.6,-1.2) {\scriptsize$S$};
			\node[ellipse,draw,line width = 1.2pt, drop shadow, fill = lightgray] (3) at (1.2,-1.2) {\scriptsize$Y(d)$};
			\node[ellipse,draw,line width = 1.2pt, drop shadow, fill = white] (4) at (2.4,-2.4) {\scriptsize$Y^*(Y(d))$};
			
			\path (1) edge (3);
			\path (2) edge (4);
			\path (3) edge (4);	
		\end{tikzpicture} \caption{\footnotesize Model $M_1$: MCAR $m$-SWIG}
	\end{subfigure}

	\begin{subfigure}{0.4\textwidth}\centering 
		\begin{tikzpicture}
			\tikzset{line width=1.5pt}
			
			\node[ellipse,draw,line width = 1.2pt, drop shadow, fill = white] (0) at (2.4,1.2) {\scriptsize$X$};
			\node[ellipse,draw,line width = 1.2pt, drop shadow, fill = white] (1) at (0,0) {\scriptsize$D$};
			\node[ellipse,draw,line width = 1.2pt, drop shadow, fill = white] (2)  at (3.6,-1.2) {\scriptsize$S$};
			\node[ellipse,draw,line width = 1.2pt, drop shadow, fill = lightgray] (3) at (1.2,-1.2) {\scriptsize$Y$};
			\node[ellipse,draw,line width = 1.2pt, drop shadow, fill = white] (4) at (2.4,-2.4) {\scriptsize$Y^*$};
			
			\path (1) edge (3);
			\path (2) edge (4);
			\path (3) edge (4);

			\path (0) edge[out=west,in=north west] (1);	
			\path (0) edge (2);	
			\path (0) edge (3);	
			
			\path(1) edge (2);
		\end{tikzpicture}\caption{\footnotesize Model $M_2$: MAR $m$-DAG}
	\end{subfigure}
	\begin{subfigure}{0.4\textwidth}\centering
		\begin{tikzpicture}
			\tikzset{line width=1.2pt,
				swig vsplit={gap=3pt,
					inner line width right=0.5pt}}
			
			\node[ellipse,draw,line width = 1.2pt, drop shadow, fill = white] (0) at (2.4,1.2) {\scriptsize$X$};
			\node[name=a1,shape=swig vsplit, drop shadow, fill = white] at (0,0) {
				\nodepart{left}{\scriptsize$D$}
				\nodepart{right}{\scriptsize$d$} } ;
			\node[ellipse,draw,line width = 1.2pt, drop shadow, fill = white] (2)  at (3.6,-1.2) {\scriptsize$S(d)$};
			\node[ellipse,draw,line width = 1.2pt, drop shadow, fill = lightgray] (3) at (1.2,-1.2) {\scriptsize$Y(d)$};
			\node[ellipse,draw,line width = 1.2pt, drop shadow, fill = white] (4) at (2.4,-2.4) {\scriptsize$Y^*(Y(d),S(d))$};
			
			\path (a1) edge (3);
			\path (2) edge (4);
			\path (3) edge (4);	
			
			to[out=350,in=170] (a1);
			
			\path (0) edge[out=west,in=north west] (a1);	
			\path (0) edge (2);	
			\path (0) edge (3);	
			
			\path(a1) edge (2);
		\end{tikzpicture} \caption{\footnotesize Model $M_2$: MAR $m$-SWIG}
	\end{subfigure}
	
	\begin{subfigure}{0.4\textwidth}\centering
		\begin{tikzpicture}
			\tikzset{line width=1.5pt}

			\node[ellipse,draw,line width = 1.2pt,dashed, drop shadow, fill = white] (-1) at (4.8,0) {\scriptsize$U$};
			\node[ellipse,draw,line width = 1.2pt, drop shadow, fill = white] (0) at (2.4,1.2) {\scriptsize$X$};
			\node[ellipse,draw,line width = 1.2pt, drop shadow, fill = white] (1) at (0,0) {\scriptsize$D$};
			\node[ellipse,draw,line width = 1.2pt, drop shadow, fill = white] (2)  at (3.6,-1.2) {\scriptsize$S$};
			\node[ellipse,draw,line width = 1.2pt, drop shadow, fill = lightgray] (3) at (1.2,-1.2) {\scriptsize$Y$};
			\node[ellipse,draw,line width = 1.2pt, drop shadow, fill = white] (4) at (2.4,-2.4) {\scriptsize$Y^*$};
			
			\path (1) edge (3);
			\path (2) edge (4);
			\path (3) edge (4);

			\path (0) edge[out=west,in=north west] (1);	
			\path (0) edge (2);	
			\path (0) edge (3);	
			
			\path[<->] (0) edge[dashed]  (-1);	
			
			\path (-1) edge (2);	
			\path (-1) edge (3);	
			\path (1) edge (2);	
			
		\end{tikzpicture} \caption{\footnotesize Model $M_3$: MNAR $m$-DAG}
	\end{subfigure}
	\begin{subfigure}{0.4\textwidth}\centering 
		\begin{tikzpicture}	\tikzset{line width=1.2pt,
				swig vsplit={gap=3pt,
					inner line width right=0.5pt}}
			
			\node[name=a1,shape=swig vsplit, drop shadow, fill = white] at (0,0) {
				\nodepart{left}{\scriptsize$D$}
				\nodepart{right}{\scriptsize$d$} } ;
			\node[ellipse,draw,line width = 1.2pt, drop shadow, fill = white] (0) at (2.4,1.2) {\scriptsize$X$};
			\node[ellipse,draw,line width = 1.2pt,dashed, drop shadow, fill = white] (-1) at (4.8,0) {\scriptsize$U$};
			\node[ellipse,draw,line width = 1.2pt, drop shadow, fill = white] (2)  at (3.6,-1.2) {\scriptsize$S(d)$};
			\node[ellipse,draw,line width = 1.2pt, drop shadow, fill = lightgray] (3) at (1.2,-1.2) {\scriptsize$Y(d)$};
			\node[ellipse,draw,line width = 1.2pt, drop shadow, fill = white] (4) at (2.4,-2.4) {\scriptsize$Y^*(Y(d),S(d))$};
			
			\path (1) edge (3);
			\path (2) edge (4);
			\path (3) edge (4);

			\path (0) edge[out=west,in=north west] (a1);	
			\path (0) edge (2);	
			\path (0) edge (3);	
			
			\path[<->] (0) edge[dashed]  (-1);
			
			\path (-1) edge (2);	
			\path (-1) edge (3);				
			\path (a1) edge (2);	
		\end{tikzpicture}\caption{\footnotesize Model $M_3$: MNAR $m$-SWIG}
	\end{subfigure}
\end{figure}
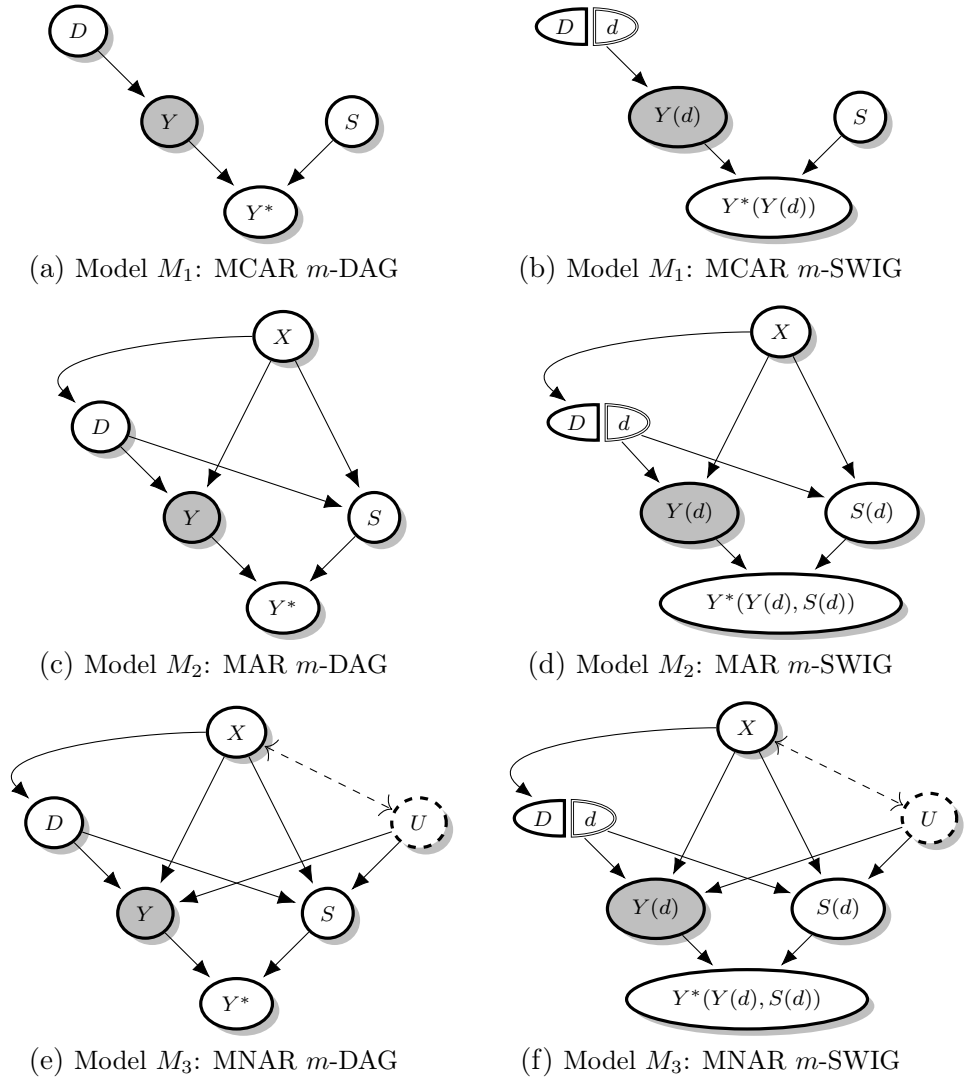

Figure \ref{fig_DS_examplesMiss1} contains the causal $m$-DAGs $\mathcal{G}$ and corresponding $m$-SWIGs $\mathcal{G}(d)$ under various selection mechanisms. One can think about variable $D$ as treatment and $Y$ as outcome in the context of an evaluation study, i.e.~a main object of interest is causal effect of this treatment on the outcome. White nodes denote observable variables in $V_O \cup V^* \cup S$. Gray nodes are partially unobserved variables in $V_m$. Unobserved variables in $U$ are denoted with dashed lines. Independent, unobserved variables that only affect single nodes are omitted as standard in DAG notation. Table \ref{tab_impliedCI_1} contains implied (conditional) independences from both $m$-DAG and $m$-SWIG for all three models. 

\begin{table}[!h]\centering \caption{Selected Implied Independence Relationships}\label{tab_impliedCI_1}
\begin{threeparttable}   \footnotesize
	\begin{tabular}{l|c|c|c}\hline \hline && \\[-.5ex] 
		Model 	& $m$-DAG 				& $m$-SWIG & Missingness \\ \hline && \\[-.5ex] 
		$M_1$ 	& $S \perp Y,D$		& $S\perp D$ & MCAR \\ 
		&  					& $Y(d)\perp D,S$ & \\ \hline && \\[-.5ex] 
		$M_2$	& $S \perp Y|X,D$	& $S(d)\perp Y(d),D|X$ & MAR \\
		&  					& $Y(d)\perp D|X$ \\ \hline && \\[-.5ex] 
		$M_3$	& $S \perp Y|X,D,U$	& $S(d)\perp Y(d)|X,U$ &MNAR \\ 
		&  					& $D\perp S(d),Y(d)|X$ \\  \hline   \hline
	\end{tabular}
	\begin{justify}\footnotesize
		Selection of implied (conditional) independences from both $m$-DAG and $m$-SWIG for each model $M_1$, $M_2$, and $M_3$. 
	\end{justify}
\end{threeparttable}
\end{table}

Model $M_1$ is the simple setup where variable $D$ causes $Y$ but $Y$ is only partially observed and no other relevant variables are linked to the system.
$S$ here causes $Y$ to be partially missing and thus links $Y$ to the observed $Y^*$. $S$ itself has only independent causes. Thus, $M_1$ implies the variable based \textit{missing-completely-at-random} condition $S \perp Y,D$. Furthermore, the treatment is as good as randomly allocated with respect to the potential outcomes. Thus, identification of an expected potential outcome is achieved by
\begin{align}
	E[Y^*|S=1,D=d] 
	&= E[Y(d)|S=1,D=d] \notag \\
	&= E[Y(d)]
\end{align}

Model $M_2$ is an extended version of $M_1$ with observed confounders. In particular, $M_2$ contains a common ancestor $X$ between treatment, selection, and outcome. Thus, selection here is \textit{missing-at-random} as $S \perp Y|X,D$. Moreover, the $m$-SWIG implies the classic (treatment) \textit{unconfoundedness} or \textit{selection-on-observables} condition. This yields identification of the (conditional) expected potential outcome:
\begin{align}
	E[Y^*|S=1,D=d,X=x]
	&= E[Y(d)|S(d)=1,D=d,X=x] \notag \\
	&= E[Y(d)|X=x]
\end{align}

Model $M_3$ introduces additional unobserved confounders $U$ for the selection step. 
A model like $M_3$ does not in general yield nonparametric point identification of the causal effect of $D$ on $Y$ due the backdoor between unobserved selection confounders $U$ and $Y$, i.e.~data is \textit{missing-not-at-random}. 
However, the corresponding $m$-SWIG can still be used to derive useful conditional independencies.  In this case, we obtain \begin{align}
	D \perp S(d),Y(d)|X \label{eq_cia_leetype1}.
\end{align} 
Condition \eqref{eq_cia_leetype1} can be used directly to derive worst-case bounds \cite{horowitz2000nonparametric} or be combined with shape restrictions to achieve narrower partial identification results \cite{zhang2003estimation,lee2009training}. For instance, one can impose a monotonicity assumption about the effect of treatment on selection $D \rightarrow S$: \begin{ass}[Selection Monotonicity]
	$P(S(1) \geq S(0)) = 1$ \label{ass_mon1}
\end{ass}  
Assumption \ref{ass_mon1} involves two hypothetical intervention variables, $S(1)$ and $S(0)$, and is thus a restriction across two SWIGs, i.e.~a \textit{cross-world} assumption. 
Alternatively, with covariates $X$, a weaker conditional monotonicity can be used:
\begin{ass}[Conditional Selection Monotonicity]
	There exist partitions of the covariate space $\mathcal{X} = \mathcal{X}_+ \cup \mathcal{X}_-$ such that $P(S(1)\geq S(0)|X\in\mathcal{X}_+) = P(S(1)\leq S(0)|X\in\mathcal{X}_-) = 1$ \label{ass_mon2}
\end{ass} 
This assumption states that, given $X$, we can predict the sign of the direction of the effect of treatment on selection, see \citeA{zhang2003estimation}, \citeA{lee2009training}, \citeA{semenova2020better}, or \citeA{heiler2022hetbounds} for more detailed discussion and Section \ref{sec_modelsandesti1} for an empirical example.\footnote{Other assumptions include, but are not limited to, mean-rank or stochastic dominance relationships for various principal strata.} Under Model $M_3$ and monotonicity, the relative proportion of always-observed, $S(0)=S(1)=1$, and compliers, $S(d)=d$, are identified for a given $x$. Thus, the missing outcomes can be bounded by trimming at the quantiles of the corresponding (best/worst-case) shares. For example, under Assumption \ref{ass_mon1}, a sharp lower bound for counterfactuals for always-observed units\footnote{Always-observed units can also be interpreted as those who have unobservables $U$ potentially related to $Y(d)$ that are sufficiently relevant such that changes in the treatment $D$ will not alter their selection status, see also \citeA{huber2015sharp} for equivalent assumptions and identification results for other principal strata beyond always-observed.} is given by 
\begin{align}
    E[Y^*|Y^*\leq q(p_0(x)),S=1,D=1,X=x] 
    &\leq E[Y|S(0)=1,D=1,X=x] \notag \\
    &= E[Y(1)|S(0)=1,S(1)=1,X=x]
\end{align}
where $q(u,x) =  \inf\{q : u \leq P(Y\leq q|S=1,D=1,X=x)\}$ denotes the conditional quantile of the observed, treated population and $p_0(x) = \tfrac{P(S=1|D=0,X=x)}{P(S=1|D=1,X=x)}$. 
An equivalent result holds for the upper bound, see Section \ref{sec_estimation} for more details and \citeA{zhang2003estimation} and \citeA{lee2009training} for first constructions of monotonicity bounds. In such problems and related, causal graphs are a first step for deriving useful conditional independencies such as \eqref{eq_cia_leetype1} and apply counterfactual adjustment even when parameters are only set-identified. 

\subsection{Testing for Selection and Attrition Bias} \label{sec_testattrition1}
In this section we use the graphical toolkit to shed light on the common practice of testing for bias and assessing necessary conditions for identification as commonly done in experimental or observational studies with missing data or attrition.
Statistical tests related to attrition or missing data in evaluation studies with (baseline) covariates $X$ are usually of one of three kinds \cite{ghanem2022testing}: (i) differential attrition, (ii) determinants of attrition, or (iii) selective attrition. Differential attrition compares attrition rate differences between control and treated units.  Determinants of attrition compares the distribution of baseline covariates for respondents and non-respondents. Selective attrition compares baseline characteristics across treated and control for given attrition status. Given covariates $X$, treatment $D$, and selection $S$, their corresponding null hypotheses translate to (i) $D \perp S|X$, (ii) $S \perp X|D$, and (iii) $X\perp D|S$.

\begin{figure}	\centering \caption{Restricted Model $M_2$ Versions} \label{fig_DS_restrictedM2}
	\begin{subfigure}{0.4\textwidth} 
		\begin{tikzpicture}
			\tikzset{line width=1.5pt}
			
			\node[ellipse,draw,line width = 1.2pt, drop shadow, fill = white] (0) at (2.4,1.2) {\scriptsize$X$};
			\node[ellipse,draw,line width = 1.2pt, drop shadow, fill = white] (1) at (0,0) {\scriptsize$D$};
			\node[ellipse,draw,line width = 1.2pt, drop shadow, fill = white] (2)  at (3.6,-1.2) {\scriptsize$S$};
			\node[ellipse,draw,line width = 1.2pt, drop shadow, fill = lightgray] (3) at (1.2,-1.2) {\scriptsize$Y$};
			\node[ellipse,draw,line width = 1.2pt, drop shadow, fill = white] (4) at (2.4,-2.4) {\scriptsize$Y^*$};
			
			\path (1) edge (3);
			\path (2) edge (4);
			\path (3) edge (4);

			\path (0) edge[out=west,in=north west] (1);	
			\path (0) edge (2);	
			\path (0) edge (3);	
			
		\end{tikzpicture}
		\caption{\footnotesize Differential Attrition}
	\end{subfigure}\begin{subfigure}{0.4\textwidth} 
		\begin{tikzpicture}
			\tikzset{line width=1.5pt}
			
			\node[ellipse,draw,line width = 1.2pt, drop shadow, fill = white] (0) at (2.4,1.2) {\scriptsize$X$};
			\node[ellipse,draw,line width = 1.2pt, drop shadow, fill = white] (1) at (0,0) {\scriptsize$D$};
			\node[ellipse,draw,line width = 1.2pt, drop shadow, fill = white] (2)  at (3.6,-1.2) {\scriptsize$S$};
			\node[ellipse,draw,line width = 1.2pt, drop shadow, fill = lightgray] (3) at (1.2,-1.2) {\scriptsize$Y$};
			\node[ellipse,draw,line width = 1.2pt, drop shadow, fill = white] (4) at (2.4,-2.4) {\scriptsize$Y^*$};
			
			\path (1) edge (3);
			\path (2) edge (4);
			\path (3) edge (4);

			\path (0) edge[out=west,in=north west] (1);	
			\path (0) edge (3);	
			
			\path(1) edge (2);
		\end{tikzpicture}
		\caption{\footnotesize Determinants of Attrition}
	\end{subfigure}	\begin{subfigure}{0.4\textwidth} 
		\begin{tikzpicture}
			\tikzset{line width=1.5pt}
			
			\node[ellipse,draw,line width = 1.2pt, drop shadow, fill = white] (0) at (2.4,1.2) {\scriptsize$X$};
			\node[ellipse,draw,line width = 1.2pt, drop shadow, fill = white] (1) at (0,0) {\scriptsize$D$};
			\node[ellipse,draw,line width = 1.2pt, drop shadow, fill = white] (2)  at (3.6,-1.2) {\scriptsize$S$};
			\node[ellipse,draw,line width = 1.2pt, drop shadow, fill = lightgray] (3) at (1.2,-1.2) {\scriptsize$Y$};
			\node[ellipse,draw,line width = 1.2pt, drop shadow, fill = white] (4) at (2.4,-2.4) {\scriptsize$Y^*$};
			
			\path (1) edge (3);
			\path (2) edge (4);
			\path (3) edge (4);

			\path (0) edge (3);	
			
			\path(1) edge (2);
		\end{tikzpicture}
		\caption{\footnotesize Selective Attrition (1)}
	\end{subfigure}	\begin{subfigure}{0.4\textwidth} 	\begin{tikzpicture}
			\tikzset{line width=1.5pt}
			
			\node[ellipse,draw,line width = 1.2pt, drop shadow, fill = white] (0) at (2.4,1.2) {\scriptsize$X$};
			\node[ellipse,draw,line width = 1.2pt, drop shadow, fill = white] (1) at (0,0) {\scriptsize$D$};
			\node[ellipse,draw,line width = 1.2pt, drop shadow, fill = white] (2)  at (3.6,-1.2) {\scriptsize$S$};
			\node[ellipse,draw,line width = 1.2pt, drop shadow, fill = lightgray] (3) at (1.2,-1.2) {\scriptsize$Y$};
			\node[ellipse,draw,line width = 1.2pt, drop shadow, fill = white] (4) at (2.4,-2.4) {\scriptsize$Y^*$};
			
			\path (1) edge (3);
			\path (2) edge (4);
			\path (3) edge (4);

			\path (0) edge (2);	
			\path (0) edge (3);	
			
		\end{tikzpicture}
		\caption{\footnotesize Selective Attrition (2)}
	\end{subfigure}	
\end{figure}
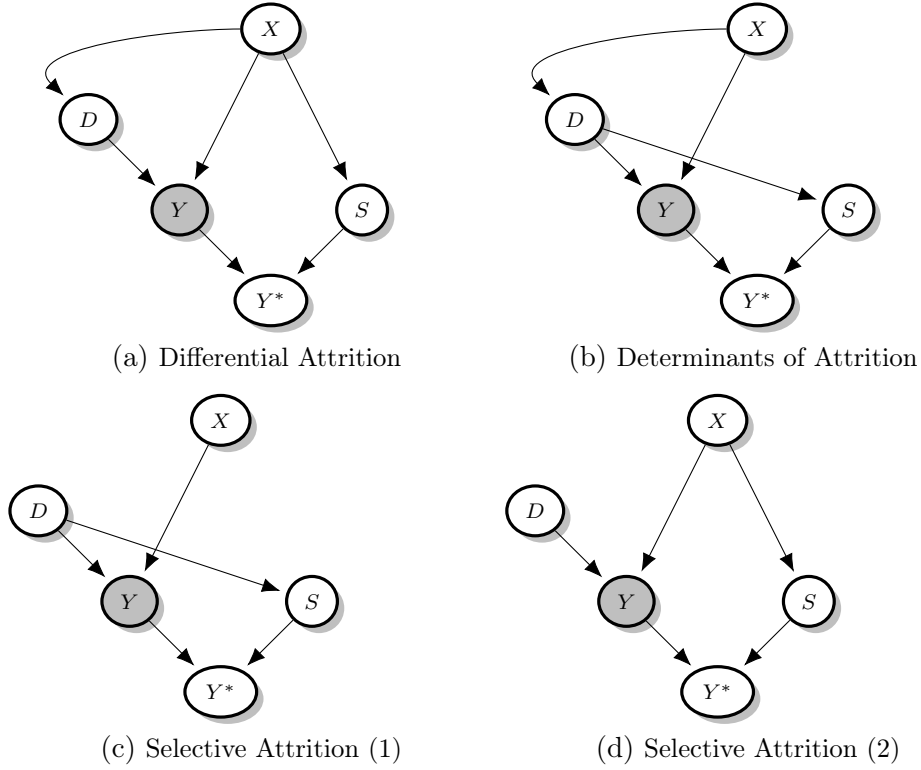

Various procedures have been suggested to empirically test these conditions, see \citeA{ghanem2022testing} for an overview. 
The utility of such tests is best understood under the missing-at-random model $M_2$. In particular, all of these strategies test \textit{implied} conditional independencies under varying causal exclusion restrictions. Figure \ref{fig_DS_restrictedM2} contains such restricted versions of model $M_2$.
Table \ref{tab_M2restrict_CIAs1} contains the corresponding exclusion restrictions and implied (conditional) independencies.      

\begin{table} \centering 
\begin{threeparttable} \caption{Comparison: Attrition Tests under Model $M_2$} \label{tab_M2restrict_CIAs1} \footnotesize
	\begin{tabular}{llc} \hline \hline 
		Type & Restrictions & Implied Independencies \\ \hline \\[-1.5ex]
		Differential Attrition & $D \nrightarrow S$ & $D \perp S | X$ \\[1ex]
		Determinants of Attrition & $X \nrightarrow S$ & $S \perp X | D$ \\[1ex]
		Selective Attrition (1) & $X \nrightarrow D$ and &  $X \perp D | S$\\
		& $X\nrightarrow S$  &  $S \perp X$ \\[1ex]
		Selective Attrition (2) & $X \nrightarrow D$ and &  $X \perp D | S$\\
		& $D\nrightarrow S$    & 	$D \perp S$	\\	\hline \hline
	\end{tabular}
	\begin{justify}\footnotesize
		This table contains the different types of attrition tests. Restriction refer to excluded edges in the context of Model $M_2$. Implied independencies are obtained by the respective restricted models using $d$-separation.
	\end{justify}
 \end{threeparttable}
\end{table}

First note that, in the case of a perfectly randomized controlled trial with respect to $D$, the exclusion restriction $X\nrightarrow D$ can credibly be imposed and thus differential attrition and determinants of attrition tests do not require conditioning on $X$ and $D$ respectively. In this case, the models in Figure \ref{fig_DS_restrictedM2} collapse to only two versions. In particular, the model behind differential attrition is equal to selective attrition (2) with implied independence $D \perp X$. Equivalently, determinants of attrition is equal to selective attrition (1) with independence $S \perp X$.\footnote{In practice, testing based on the conditional independencies can still be preferred due to an increase in the number of available restrictions. 
In either case, if the model is not rejected, identification is more likely to be credible in the randomized experiment despite missing data.}

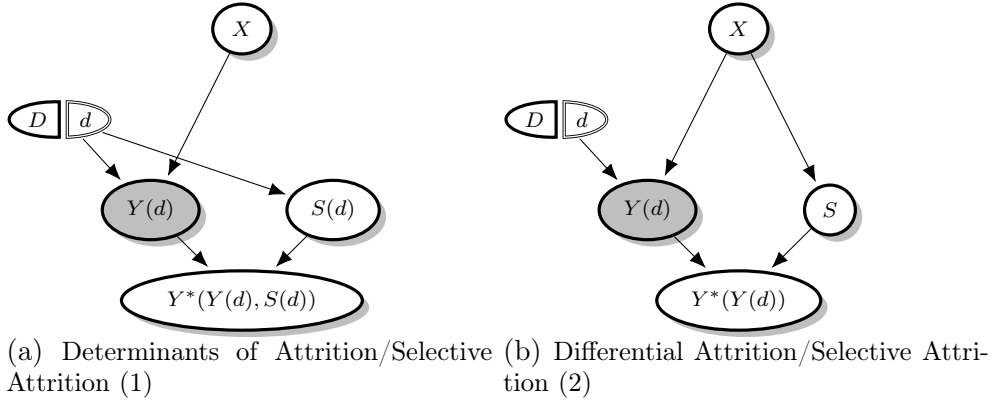
\begin{figure}	\centering \caption{Randomized Experiment Submodel SWIGs} \label{fig_SWIGs_M2withRandomization}
	\begin{subfigure}{0.4\textwidth} 
		\begin{tikzpicture}
			\tikzset{line width=1.2pt,
				swig vsplit={gap=3pt,
					inner line width right=0.5pt}}
			
			\node[ellipse,draw,line width = 1.2pt, drop shadow, fill = white] (0) at (2.4,1.2) {\scriptsize$X$};
			\node[name=a1,shape=swig vsplit, drop shadow, fill = white] at (0,0) {
				\nodepart{left}{\scriptsize$D$}
				\nodepart{right}{\scriptsize$d$} } ;
			\node[ellipse,draw,line width = 1.2pt, drop shadow, fill = white] (2)  at (3.6,-1.2) {\scriptsize$S(d)$};
			\node[ellipse,draw,line width = 1.2pt, drop shadow, fill = lightgray] (3) at (1.2,-1.2) {\scriptsize$Y(d)$};
			\node[ellipse,draw,line width = 1.2pt, drop shadow, fill = white] (4) at (2.4,-2.4) {\scriptsize$Y^*(Y(d),S(d))$};
			
			\path (a1) edge (3);
			\path (2) edge (4);
			\path (3) edge (4);	
			
			to[out=350,in=170] (a1);
			
			\path (0) edge (3);	
			
			\path(a1) edge (2);
		\end{tikzpicture}
		\caption{\footnotesize Determinants of Attrition/Selective Attrition (1)}
	\end{subfigure}	\begin{subfigure}{0.4\textwidth} \begin{tikzpicture}
			\tikzset{line width=1.2pt,
				swig vsplit={gap=3pt,
					inner line width right=0.5pt}}
			
			\node[ellipse,draw,line width = 1.2pt, drop shadow, fill = white] (0) at (2.4,1.2) {\scriptsize$X$};
			\node[name=a1,shape=swig vsplit, drop shadow, fill = white] at (0,0) {
				\nodepart{left}{\scriptsize$D$}
				\nodepart{right}{\scriptsize$d$} } ;
			\node[ellipse,draw,line width = 1.2pt, drop shadow, fill = white] (2)  at (3.6,-1.2) {\scriptsize$S$};
			\node[ellipse,draw,line width = 1.2pt, drop shadow, fill = lightgray] (3) at (1.2,-1.2) {\scriptsize$Y(d)$};
			\node[ellipse,draw,line width = 1.2pt, drop shadow, fill = white] (4) at (2.4,-2.4) {\scriptsize$Y^*(Y(d))$};
			
			\path (a1) edge (3);
			\path (2) edge (4);
			\path (3) edge (4);	
			
			to[out=350,in=170] (a1);
			
			\path (0) edge (2);	
			\path (0) edge (3);	
			
		\end{tikzpicture}
		\caption{\footnotesize Differential Attrition/Selective Attrition (2)}
	\end{subfigure}	
\end{figure}

Consider now the two $m$-SWIGs for the Model $M_2$ cases with the randomization exclusion $X\nrightarrow D$ in Figure \ref{fig_SWIGs_M2withRandomization}.  
The determinants of attrition/selective attrition (1) model then implies that $Y(d) \perp S,D$ (with or without conditioning on $X$) and thus
\begin{align}
	E[Y^*|S=1,D=d] &= E[Y(d)|S(d)=1,D=d] \notag \\
	&= E[Y(d)]
\end{align} 
and equivalently for the conditional mean $E[Y(d)|X]$. Hence, the selected population yields a representative potential outcome/treatment effect because potential selection is not caused by covariates $X$. Note that, even when unconditional estimands are of interest, $X$ is a neutral control and could still be useful to increase efficiency when included in the outcome process model \cite{cinelli2020crash}. 
In contrast, in the differential attrition/selective attrition (2) case, the restricted model only implies that $Y(d) \perp D$ (with or without conditioning on $X$) but only $Y(d)\perp S|X$ and hence 
\begin{align}
	E[Y^*|S=1,D=d,X] &=E[Y(d)|S=1,X] \notag \\
	&= E[Y(d)|X].
\end{align}
Thus, the potential outcome/treatment effect is only representative for the population after conditioning on $X$ due to these variables causing both selection and (potential) outcome. Thus, even in large samples, non-rejection of differential attrition is not sufficient to ignore covariates without additional assumptions even under $M_2$. 

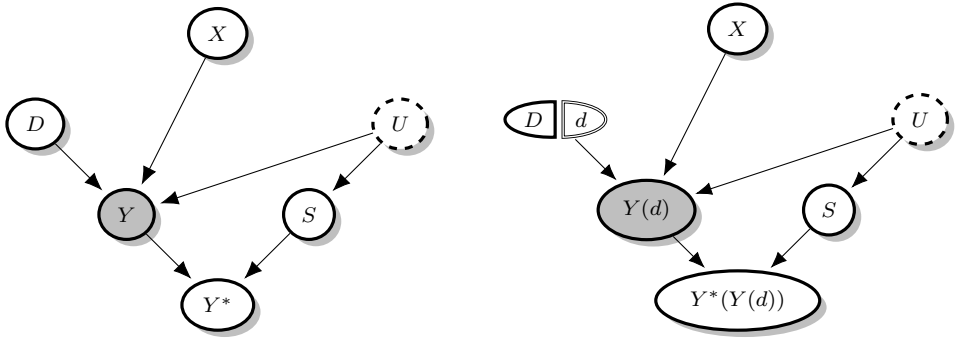
\begin{figure}[!h] \centering \caption{Restricted Model $M_3$ DAG and SWIG} \label{fig_DS_restrictedM3}
	\begin{subfigure}{0.4\textwidth}
		\begin{tikzpicture}
			\tikzset{line width=1.5pt}

			\node[ellipse,draw,line width = 1.2pt,dashed, drop shadow, fill = white] (-1) at (4.8,0) {\scriptsize$U$};
			\node[ellipse,draw,line width = 1.2pt, drop shadow, fill = white] (0) at (2.4,1.2) {\scriptsize$X$};
			\node[ellipse,draw,line width = 1.2pt, drop shadow, fill = white] (1) at (0,0) {\scriptsize$D$};
			\node[ellipse,draw,line width = 1.2pt, drop shadow, fill = white] (2)  at (3.6,-1.2) {\scriptsize$S$};
			\node[ellipse,draw,line width = 1.2pt, drop shadow, fill = lightgray] (3) at (1.2,-1.2) {\scriptsize$Y$};
			\node[ellipse,draw,line width = 1.2pt, drop shadow, fill = white] (4) at (2.4,-2.4) {\scriptsize$Y^*$};
			
			\path (1) edge (3);
			\path (2) edge (4);
			\path (3) edge (4);

			\path (0) edge (3);	
			
			
			\path (-1) edge (2);	
			\path (-1) edge (3);	
			
		\end{tikzpicture}
	\end{subfigure}
	\begin{subfigure}{0.4\textwidth}
		\begin{tikzpicture}	\tikzset{line width=1.2pt,
				swig vsplit={gap=3pt,
					inner line width right=0.5pt}}
			
			\node[name=a1,shape=swig vsplit, drop shadow, fill = white] at (0,0) {
				\nodepart{left}{\scriptsize$D$}
				\nodepart{right}{\scriptsize$d$} } ;
			\node[ellipse,draw,line width = 1.2pt, drop shadow, fill = white] (0) at (2.4,1.2) {\scriptsize$X$};
			\node[ellipse,draw,line width = 1.2pt,dashed, drop shadow, fill = white] (-1) at (4.8,0) {\scriptsize$U$};
			\node[ellipse,draw,line width = 1.2pt, drop shadow, fill = white] (2)  at (3.6,-1.2) {\scriptsize$S$};
			\node[ellipse,draw,line width = 1.2pt, drop shadow, fill = lightgray] (3) at (1.2,-1.2) {\scriptsize$Y(d)$};
			\node[ellipse,draw,line width = 1.2pt, drop shadow, fill = white] (4) at (2.4,-2.4) {\scriptsize$Y^*(Y(d))$};
			
			\path (1) edge (3);
			\path (2) edge (4);
			\path (3) edge (4);

			\path (0) edge (3);	
			
			
			\path (-1) edge (2);	
			\path (-1) edge (3);				
		\end{tikzpicture}
	\end{subfigure}
\end{figure}

A more severe problem with all these tests is revealed under a deviation of the truth from model $M_2$. In the following we briefly discuss why, even in the most favorable case where conditional independencies are both sufficient and necessary for the model space (and not just necessary as before), non-rejection of the conditional independencies is neither sufficient nor necessary for identification. This most favorable case is a simplification and without loss of generality. Assume now that the truth is generated by model $M_3$ as in Figure \ref{fig_DS_examplesMiss1} and that both determinants of attrition and differential attrition have been tested. In the case of non-rejection of both hypotheses and equivalence of model and hypothesis space, DAG and SWIG for the underlying restricted $M_3$ are given in Figure \ref{fig_DS_restrictedM3}.  

The restricted model implies both $D \perp S$, $X \perp S$. However, in contrast to $M_2$, we no longer have that $Y(d) \perp S | X$. Thus $E[Y^*|S=1,D=d,X] \neq E[Y(d)|X]$ in general and identification fails.

\begin{figure} \centering \caption{Model that Rejects all Attrition Test}\label{fig_DAG_rejectall}
	\begin{subfigure}{0.4\textwidth}
		\begin{tikzpicture}
			\tikzset{line width=1.5pt}

			\node[ellipse,draw,line width = 1.2pt,dashed, drop shadow, fill = white] (-1) at (4.8,0) {\scriptsize$U$};
			\node[ellipse,draw,line width = 1.2pt, drop shadow, fill = white] (0) at (2.4,1.2) {\scriptsize$X$};
			\node[ellipse,draw,line width = 1.2pt, drop shadow, fill = white] (1) at (0,0) {\scriptsize$D$};
			\node[ellipse,draw,line width = 1.2pt, drop shadow, fill = white] (2)  at (3.6,-1.2) {\scriptsize$S$};
			\node[ellipse,draw,line width = 1.2pt, drop shadow, fill = lightgray] (3) at (1.2,-1.2) {\scriptsize$Y$};
			\node[ellipse,draw,line width = 1.2pt, drop shadow, fill = white] (4) at (2.4,-2.4) {\scriptsize$Y^*$};
			
			\path (2) edge (4);
			\path (3) edge (4);

			
			\path (0) edge (-1);	
			\path (1) edge (-1);	
			
			\path (-1) edge (2);	
			
		\end{tikzpicture}
	\end{subfigure}	
\end{figure}
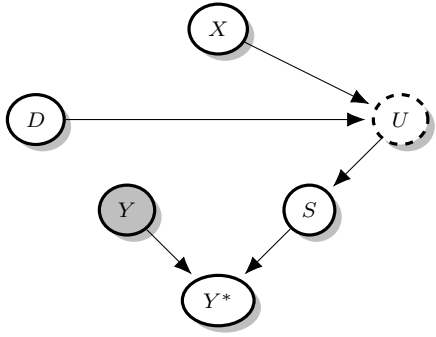

On the other hand, rejection of all the conditional independencies is also not necessary for identification. Consider the stylized example in Figure \ref{fig_DAG_rejectall}.
In this model, while neither $D \perp S$ nor $X\perp S$ holds, the selected sample is clearly representative as it corresponds to a missing-completely-at-random $Y \perp S$ scenario with $Y^*\perp D|S$ and thus $E[Y^*|D=1,S=1] = E[Y]$. 
Of course, this is a stylized example (that could potentially be rejected by testing $X\perp D$ and $X \perp Y^*|S$). However, it illustrates that, when the goal is identification of an expected potential outcome/treatment effect, such proxy tests can lead to false conclusions under varying causal structures. Therefore, it seems warranted to take all these tests only as indication and thus reporting estimates from different specifications under increasingly weak assumptions still remains important.

\subsection{Application to Panel Data and Field Experiments}
\begin{figure} \centering \caption{Panel Data Model $M_4$} \label{fig_DS_panelM4} 
\begin{subfigure}{0.3\textwidth}  
	\begin{tikzpicture}
		\tikzset{line width=1.5pt}

		\node[ellipse,draw,line width = 1.2pt, drop shadow, fill = white] (1) at (0,0) {\scriptsize$D$};
		\node[ellipse,draw,line width = 1.2pt, drop shadow, fill = white] (2)  at (1.2,-1.2) {\scriptsize$Y_0$};
		\node[ellipse,draw,line width = 1.2pt, drop shadow, fill = lightgray] (3) at (2.4,-1.2) {\scriptsize$Y_1$};
		\node[ellipse,draw,line width = 1.2pt, drop shadow, fill = white] (4) at (3.6,0) {\scriptsize$S$};
		
		\node[ellipse,draw,line width = 1.2pt,dashed, drop shadow, fill = white] (5) at (1.2,-2.4) {\scriptsize$U_0$};
		\node[ellipse,draw,line width = 1.2pt,dashed, drop shadow, fill = white] (6) at (2.4,-2.4) {\scriptsize$U_1$};
		\node[ellipse,draw,line width = 1.2pt,dashed, drop shadow, fill = white] (7) at (3.6,-2.4) {\scriptsize$V$};

		\path (1) edge (4);	
		\path (1) edge[out=south east,in=north] (3);

		\path (5) edge (2);	
		\path (6) edge (3);	
		\path (7) edge (4);	
		
		\path (5) edge[out=south,in=south,dashed,<->] (6);
		\path (6) edge[out=south,in=south,dashed,<->] (7);
		\path (5) edge[out=south,in=south,dashed,<->] (7);
		
	\end{tikzpicture}
	\caption{\footnotesize  DAG w/o Exclusion} \label{fig_DS_panelM4a}
\end{subfigure}	
\begin{subfigure}{0.3\textwidth}
	\begin{tikzpicture}
		\tikzset{line width=1.5pt}

		\node[ellipse,draw,line width = 1.2pt, drop shadow, fill = white] (1) at (0,0) {\scriptsize$D$};
		\node[ellipse,draw,line width = 1.2pt, drop shadow, fill = white] (2)  at (1.2,-1.2) {\scriptsize$Y_0$};
		\node[ellipse,draw,line width = 1.2pt, drop shadow, fill = lightgray] (3) at (2.4,-1.2) {\scriptsize$Y_1$};
		\node[ellipse,draw,line width = 1.2pt, drop shadow, fill = white] (4) at (3.6,0) {\scriptsize$S$};
		
		\node[ellipse,draw,line width = 1.2pt,dashed, drop shadow, fill = white] (5) at (1.2,-2.4) {\scriptsize$U_0$};
		\node[ellipse,draw,line width = 1.2pt,dashed, drop shadow, fill = white] (6) at (2.4,-2.4) {\scriptsize$U_1$};
		\node[ellipse,draw,line width = 1.2pt,dashed, drop shadow, fill = white] (7) at (3.6,-2.4) {\scriptsize$V$};

		\path (1) edge (4);	
		\path (1) edge[out=south east,in=north] (3);

		\path (5) edge (2);	
		\path (6) edge (3);	
		\path (7) edge (4);	
		
		\path (5) edge[out=south,in=south,dashed,<->] (6);
		\path (6) edge[out=south,in=south,dashed,-, color=white] (7);
		\path (5) edge[out=south,in=south,dashed,-, color=white] (7);
		
	\end{tikzpicture}
	\caption{\footnotesize  DAG w/Exclusion I} \label{fig_DS_panelM4b}
\end{subfigure}		\begin{subfigure}{0.3\textwidth}
	\begin{tikzpicture}
		\tikzset{line width=1.2pt,
			swig vsplit={gap=3pt,
				inner line width right=0.5pt}}
		
		\node[name=1,shape=swig vsplit, drop shadow, fill = white] at (0,0) {
			\nodepart{left}{\scriptsize$D$}
			\nodepart{right}{\scriptsize$d$} } ;

		\node[ellipse,draw,line width = 1.2pt, drop shadow, fill = white] (2)  at (1.2,-1.2) {\scriptsize$Y_0$};
		\node[ellipse,draw,line width = 1.2pt, drop shadow, fill = lightgray] (3) at (2.4,-1.2) {\scriptsize$Y_1(d)$};
		\node[ellipse,draw,line width = 1.2pt, drop shadow, fill = white] (4) at (3.6,0) {\scriptsize$S(d)$};
		
		\node[ellipse,draw,line width = 1.2pt,dashed, drop shadow, fill = white] (5) at (1.2,-2.4) {\scriptsize$U_0$};
		\node[ellipse,draw,line width = 1.2pt,dashed, drop shadow, fill = white] (6) at (2.4,-2.4) {\scriptsize$U_1$};
		\node[ellipse,draw,line width = 1.2pt,dashed, drop shadow, fill = white] (7) at (3.6,-2.4) {\scriptsize$V$};

		\path (1) edge (4);	
		\path (1) edge[out=south east,in=north] (3);

		\path (5) edge (2);	
		\path (6) edge (3);	
		\path (7) edge (4);	
		
		\path (5) edge[out=south,in=south,dashed,<->] (6);
		\path (6) edge[out=south,in=south,dashed,-, color=white] (7);
		\path (5) edge[out=south,in=south,dashed,-, color=white] (7);
		
	\end{tikzpicture}
	\caption{\footnotesize  SWIG w/Exclusion I} \label{fig_DS_panelM4c}
\end{subfigure}
\begin{subfigure}{0.3\textwidth}  
	\begin{tikzpicture}
		\tikzset{line width=1.5pt}

		\node[ellipse,draw,line width = 1.2pt, drop shadow, fill = white] (1) at (0,0) {\scriptsize$D$};
		\node[ellipse,draw,line width = 1.2pt, drop shadow, fill = white] (2)  at (1.2,-1.2) {\scriptsize$Y_0$};
		\node[ellipse,draw,line width = 1.2pt, drop shadow, fill = lightgray] (3) at (2.4,-1.2) {\scriptsize$Y_1$};
		\node[ellipse,draw,line width = 1.2pt, drop shadow, fill = white] (4) at (3.6,0) {\scriptsize$S$};
		
		\node[ellipse,draw,line width = 1.2pt,dashed, drop shadow, fill = white] (5) at (1.2,-2.4) {\scriptsize$U_0$};
		\node[ellipse,draw,line width = 1.2pt,dashed, drop shadow, fill = white] (6) at (2.4,-2.4) {\scriptsize$U_1$};
		\node[ellipse,draw,line width = 1.2pt,dashed, drop shadow, fill = white] (7) at (3.6,-2.4) {\scriptsize$V$};

		\path (1) edge[out=south east,in=north] (3);

		\path (5) edge (2);	
		\path (6) edge (3);	
		\path (7) edge (4);	
		
		\path (5) edge[out=south,in=south,dashed,<->] (6);
		\path (6) edge[out=south,in=south,dashed,<->] (7);
		\path (5) edge[out=south,in=south,dashed,<->] (7);
		
	\end{tikzpicture}
	\caption{\footnotesize  DAG w/ Exclusion II} \label{fig_DS_panelM4d}
\end{subfigure}		\begin{subfigure}{0.3\textwidth}
	\begin{tikzpicture}
		\tikzset{line width=1.2pt,
			swig vsplit={gap=3pt,
				inner line width right=0.5pt}}
		
		\node[name=1,shape=swig vsplit, drop shadow, fill = white] at (0,0) {
			\nodepart{left}{\scriptsize$D$}
			\nodepart{right}{\scriptsize$d$} } ;

		\node[ellipse,draw,line width = 1.2pt, drop shadow, fill = white] (2)  at (1.2,-1.2) {\scriptsize$Y_0$};
		\node[ellipse,draw,line width = 1.2pt, drop shadow, fill = lightgray] (3) at (2.4,-1.2) {\scriptsize$Y_1(d)$};
		\node[ellipse,draw,line width = 1.2pt, drop shadow, fill = white] (4) at (3.6,0) {\scriptsize$S$};
		
		\node[ellipse,draw,line width = 1.2pt,dashed, drop shadow, fill = white] (5) at (1.2,-2.4) {\scriptsize$U_0$};
		\node[ellipse,draw,line width = 1.2pt,dashed, drop shadow, fill = white] (6) at (2.4,-2.4) {\scriptsize$U_1$};
		\node[ellipse,draw,line width = 1.2pt,dashed, drop shadow, fill = white] (7) at (3.6,-2.4) {\scriptsize$V$};

		\path (1) edge[out=south east,in=north] (3);

		\path (5) edge (2);	
		\path (6) edge (3);	
		\path (7) edge (4);	
		
		\path (5) edge[out=south,in=south,dashed,<->] (6);
		\path (6) edge[out=south,in=south,dashed,<->] (7);
		\path (5) edge[out=south,in=south,dashed,<->] (7);
		
	\end{tikzpicture}
	\caption{\footnotesize  SWIG w/Exclusion II} \label{fig_DS_panelM4e}
\end{subfigure}
\end{figure}
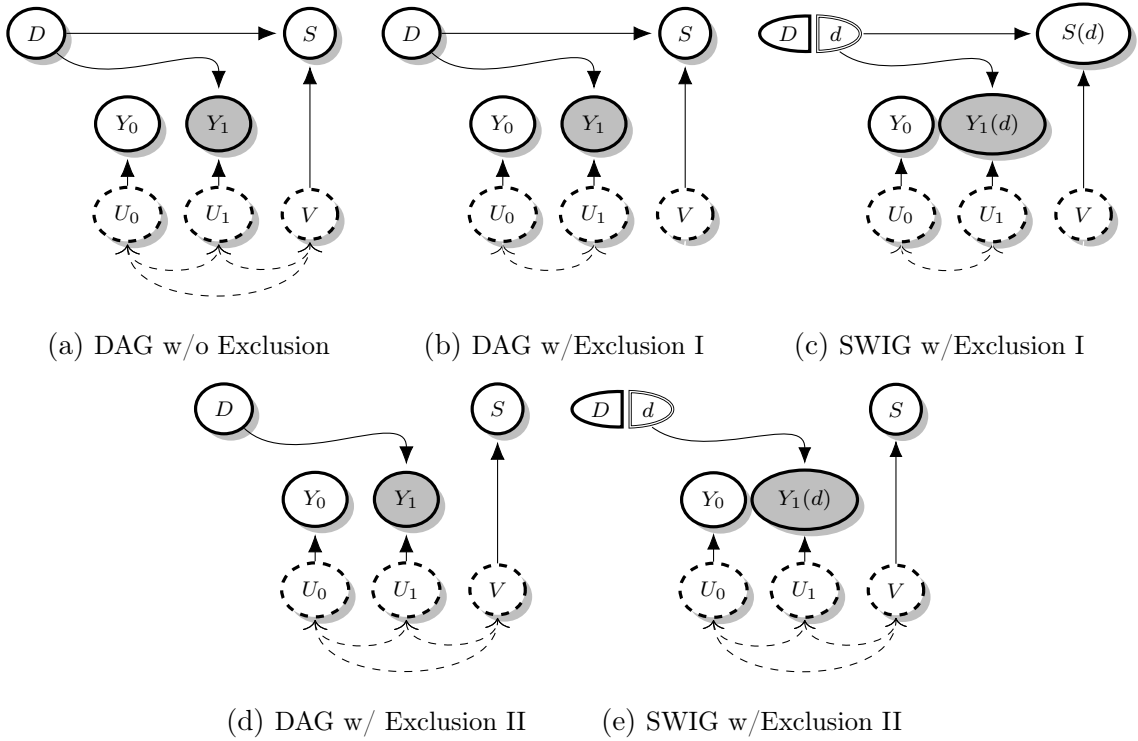

We now apply the methodology to the evaluation of randomized field experiments with attrition. In particular, we show how to construct sharp model specification tests for internal validity. We consider the case of a two-period non-separable panel model where only part of the population receives treatment and some outcomes are missing in the second period as in \citeA{ghanem2022testing} (with slightly modified notation). Covariates are omitted for simplicity. In particular, we study the following setup: There are two periods $t=0,1$ and a treatment $D = (D_0,D_1)$ that is either $(0,0)$ or $(0,1)$ characterizing control and treated respectively. Note that all of the following discussion also applies to the case of a non-binary discrete or continuous treatment. The outcome $Y_t$ in a period $t$ is assumed to be determined by \begin{align}
Y_t = f_y(D_t,U_t) \label{eq_panel_struct_Y}
\end{align}
where $f_y(\cdot)$ is a structural function and $U_t$ are unobservables containing e.g.~time-invariant (fixed) and time-varying characteristics. Selection of responses in period $t=1$ are determined by \begin{align}
S = f_s(D,V) \label{eq_panel_struct_S}
\end{align} 
where $f_s(\cdot)$ is a structural function and $V$ are unobservables causing response status. In the non-separable panel data literature, it is very common to essentially adopt either a structural or a semi-structural equation model approach with independence conditions \cite{altonji2005crossregressors,huber2012identification,chernozhukov2013average,ghanem2022testing}. For example \citeA{ghanem2022testing} impose structural equations \eqref{eq_panel_struct_Y} and \eqref{eq_panel_struct_S} but no structural equation for the treatment assignment of $D$. Instead, they impose the independence assumption on the unobservables $U_0,U_1,V \perp D$ as ensured by the experimental randomization. However, such treatment assignment is the step about which we often have most process knowledge and thus could credibly impose exclusion restrictions directly. The utility of using ``only'' a conditional independence for the treatment assignment in the context of a field experiment is unclear given that a structural paradigm has already been adopted for \eqref{eq_panel_struct_Y} and \eqref{eq_panel_struct_S}. 
From a practical perspective, it is difficult to conceive a scenario where $U_0,U_1,V \perp D$ can be reasoned for without structural knowledge about the assignment mechanism; in particular without an assignment that excludes causal links between unobservables related to outcome and selection to the treatment assignment relevant variable(s). Formally, such a structural randomized treatment assignment would be encoded as \begin{align}
D = f_d(\varepsilon_D) \label{eq_panel_struct_D}
\end{align}
where $f_d(\cdot)$ is a structural function and no edges are allowed between $\varepsilon_D$ and any other variables in the system as ensured by independent randomization, e.g.~via coin flip. Given these exclusions, we can treat $D$ and $\varepsilon_D$ as synonymous in what follows. Again, it is not clear how to construct a credible example where $U_0,U_1,V \perp D$ and \eqref{eq_panel_struct_Y} and \eqref{eq_panel_struct_S} are assumed but not \eqref{eq_panel_struct_D}. Taking all these equations together, yields a fully nonparametric structural causal model with corresponding DAG as depicted in Figure \ref{fig_DS_panelM4a}.  

The edges between $U_0$ and $U_1$ could e.g.~be due to the presence of a classic fixed effect or other forms of time-invariant causes. 
The additional exclusion $U_0,U_1,V \nrightarrow D$ then implies the independent assignment condition with respect to the treatment $U_0,U_1,V \perp D$ as in \citeA{ghanem2022testing}.
Figure \ref{fig_DS_panelM4a} makes it transparent that, in the presence of missingness, random assignment does not ensure that an evaluation is representative. In particular, $S$ is a colliding node on the path between $D\rightarrow S \leftarrow V \leftrightarrow U\rightarrow Y_1$ and thus conditioning on observing the outcome in period $t=1$  (responders) opens up a back-door that biases the causal validity of a comparison of treated and control units. \citeA{ghanem2022testing} suggest to derive sharp testable conditions that are necesseary (i.e.~implied) conditions for validity of a causal evaluation for different subpopulations. In particular, they are interested in internal validity of either the full population (i.e.~unconditional) or the respondent ($S=1$) subpopulation.

For the full population, they show that assuming  $(U_0,U_1)\perp S|D$ (``IV-P'' assumption) implies the testable sharp restriction $Y_0 \perp D,S$. Given model $M_4$, imposing a (structural) exclusion that implies $(U_0,U_1)\perp S|D$ in general seems only conceivable by eliminating the bi-directed edges between $(U_0,U_1)$ and $V$, i.e.~assuming that (i) there are no common causes of unobservables causing the outcome and unobservables causing the selection and (ii) these unobservables are not causes of each other. This implies a missing completely at random scenario for $Y_1$. The corresponding restricted model is depicted in Figure \ref{fig_DS_panelM4b}. By $d$-separation it follows immediately that the implied conditional independencies from the graph are: $Y_1 \perp S|D$ and $Y_0\perp D,S$. The first condition is not testable as we only observe the distribution of $Y_1$ conditional on being responder $S=1$. The second is the same sharp testable condition $Y_0\perp D,S$ as derived by \citeA{ghanem2022testing}. Hence the structural and semi-structural approach yields the identical conclusion.

This is different when considering the responding population $S=1$. Here \citeA{ghanem2022testing} show that $(U_0,U_1)\perp D|S$ (``IV-R'' assumption) allows for identification of the causal effect of responders. IV-R also implies the testable sharp restriction $Y_0 \perp D|S$. Under Model $M_4$, however, there are only two credible exclusion restrictions conceivable that could agree with the IV-R assumption: First, removing the causal links between unobservables $(U_0,U_1)$ and $V$ completely, i.e.~imposing exactly the same exclusion restriction as in the previous example in Figure \ref{fig_DS_panelM4b}. The corresponding $m$-SWIG in Figure \ref{fig_DS_panelM4c} then reveals that $Y(d) \perp D,S$ and thus the effect on the responders is equal to the effect on the general population. Second, assuming that there is no causal effect of treatment on selection $D\nrightarrow S$, i.e.~$D$ does not enter equation \eqref{eq_panel_struct_S}, also implies IV-R, see Figure \ref{fig_DS_panelM4d}. In this case, SWIG \ref{fig_DS_panelM4e} clearly implies that $Y(d)\perp D|S$ and thus, the average treatment effect on the selected is identified by comparing treated and control units in the observed population. However, in this case \ref{fig_DS_panelM4d} implies the sharp set of testable restrictions $D\perp Y_0$ and $D\perp S$. These two together do imply the \citeA{ghanem2022testing} restriction $Y_0 \perp D|S$, but are generally stronger. Thus, in any case, we obtain more power to reject a model that would suggest internal validity for the responders compared to the semi-structural approach.\footnote{\citeA{ghanem2022testing}, Proposition 3(ii) also shows that $D \perp S$ together with monotonicity assumption \ref{ass_mon1} imply IV-R. Such shape restrictions are complementary to the exclusion based independencies discussed in this section.} 


Hypothetically, one could also construct a scenario where the primitive independence assumption $U_0,U_1,V \perp D$ holds without imposing exclusion restrictions with regards to the treatment assignment as in Model $M_4$ and where the dependence generated by conditioning on $S$ is offset through (a multitude of) dependencies that cancel with respect to the conditional distribution $Y_0|D,S$. We consider such scenarios as mostly an academic thought exercise and essentially irrelevant for empirical practice. 
In addition, such \textit{accidental independencies} are increasingly hard to justify in more complicated scenarios with e.g.~repeated follow-ups, observed confounders, and/or dynamic dependencies as these are part of the general joint probabilistic dependence structure. Instead, an approach that encodes the structure of such complications and relies on derived independencies circumvents this issue and is transparent about the source and exclusion restrictions behind the necessary conditions. 
As in the IV-R example above, this will generally lead to an equal or larger amount of necessary restrictions compared to the purely conditional independence based approach, reducing the frequency of false non-rejection of models that produce biased causal comparisons. Extensions to more complicated scenarios e.g.~with additional time periods and observed confounders follow analogously.

\section{Estimation} \label{sec_estimation}
\subsection{From Counterfactual Adjustment and Weighting to Moment}
In this section we outline how to construct estimators based on the counterfactual adjustment formula and weighting for a given $m$-SWIG. We focus on augmented inverse probability weighting (AIPW) type moment based estimators. Due to their orthogonality property, they can be used in conjunction with flexible machine learning or other non/semiparametric estimators for their required nuisance functions $\eta$ under appropriate sample splitting/cross-fitting \cite{chernozhukov2018double}. In particular, propensity scores, selection probabilities, and potential outcome means can be estimated via modern methods such as deep neural nets, random forests, high-dimensional sparse regression, boosting or else. Valid inference can then be conducted under relatively weak assumptions. We also sketch how to use the underlying signals in the context of partial identification and for heterogeneity analysis. 

The steps are as follows: Given a model, first an $m$-SWIG is used to derive independencies. Second, the counterfactual adjustment formula is applied. Third, a moment based on the (weighted) conditional expectation functions is created. If a conditional parameter is of interest, conditional expectation functions are reweighted. Fourth, inverse probability weighted bias-correction terms are added. 
For any auxiliary derivations consider Appendix \ref{sec_App_moments1}. 
For simplifying derivations, we assume that $X$ is discrete in what follows. This is without loss of generality and the resulting moment functions will be identical for the continuous or mixed case.
In this case counterfactual adjustment as in Definition \ref{def_cfa_ipw1} can be rewritten as  \begin{align}
E[Y(d)] = \sum_{x}E[Y|X=x,D=d]P(X=x).
\end{align} 

\subsection{Example 1: Average Treatment Effect in the Generic Case}
We are interested in the average treatment effect $E[Y(1)-Y(0)]$ in a model without missing data that obeys assumption $Y(d)\perp D|X$ and we observe independent $W = (Y,D,X)$. 
Thus, using adjustment and weighting, we can construct a debiased signal as \begin{align}
\psi_d(W,\eta) = \frac{(Y-E[Y|X,D=d])}{P(D=d|X)}\mathbbm{1}(D=d) + E[Y|X,D=d] - E[Y(d)]
\end{align}
which yields the classical efficient influence function \cite{hahn1998role} for the ATE  \begin{align}
\psi(W,\eta) &= \psi_1(W,\eta) - \psi_0(W,\eta) \notag \\
&= \frac{(Y-E[Y|X,D=1])}{P(D=1|X)}D - \frac{(Y-E[Y|X,D=0])}{P(D=0|X)}(1-D) \notag  \\ &\ +  E[Y|X,D=1] - E[Y|X,D=0] -  E[Y(1)-Y(0)].
\end{align}

\subsection{Example 2: Average Treatment Effect on the Treated in Model $M_2$}
Now we consider the $m$-SWIG of $M_2$, i.e.~the setup with missing outcome data and selection on observables. We are interested in the average treatment effect on the treated $E[Y(1)-Y(0)|D=1]$ and we observe $(Y^*,S,D,X)$.  Applying the counterfactual adjustment formula to any $d,d'$ yields \begin{align}
E[Y(d)|D=d'] &= \sum_{x}E[Y(d)|D=d',X=x]P(X=x|D=d') \notag \\
&= \sum_{x}E[Y(d)|D=d',X=x]\frac{P(D=d'|X=x)}{P(D=d')}P(X=x)\notag  \\
&= \sum_{x}E[Y|S=1,D=d,X=x]\frac{P(D=d'|X=x)}{P(D=d')}P(X=x) 
\end{align}
where the last line follows from $E[Y(d)|D=d',X=x] = E[Y(d)|D=d,S(d)=1,X=x]$. Using this for two cases $d=1,d'=1$ and $d=0,d'=1$ yields \begin{align}
E[Y(1)|D=1] &= \sum_{x}E[Y|S=1,D=1,X=x]\frac{P(D=1|X=x)}{P(D=1)}P(X=x) \\
E[Y(0)|D=1] &= \sum_{x}E[Y|S=1,D=0,X=x]\frac{P(D=1|X=x)}{P(D=1)}P(X=x)
\end{align}
For simplicity, we assume that $P(D=1|X)$ is known as in the case of a (stratified) randomized experiment. If this is estimated, an additional correction is recommended \cite{newey1994asymptotic,chernozhukov2018double}. 
We can then construct a debiased moment for $E[Y(1)|D=1]$ as \begin{align}
\psi_{11}(W,\eta) &=  \frac{(Y-E[Y|S=1,D=1,X])}{P(SD=1|X)}SD\frac{P(D=1|X)}{P(D=1)} \notag  \\  &\quad +  E[Y|S=1,D=1,X]\frac{P(D=1|X)}{P(D=1)} - E[Y(1)|D=1] 
\end{align}
and for $E[Y(0)|D=1]$ as
\begin{align}	\psi_{01}(W,\eta) &=  \frac{(Y-E[Y|S=1,D=0,X])}{P(S(1-D)=1|X)}S(1-D)\frac{P(D=1|X)}{P(D=1)} \notag  \\  +  E[Y|S=1&,D=0,X]\frac{P(D=1|X)}{P(D=1)}\frac{SD}{P(SD=1)} - E[Y(0)|D=1]\frac{SD}{P(SD=1)} 
\end{align}
The moment for the ATT is then again equal to the difference between the two. 
These two moment functions are related but different compared to the efficient influence functions for the counterfactuals without missing data \cite{farrell2015robust}. In particular, without missing outcome variables, the $E[Y(1)|D=1]$ moment function simplifies as it is identified without conditioning on covariates. 

\subsection{Example 3: Zhang-Rubin-Lee Bounds without exclusion}
We are interested in $E[Y(1)-Y(0)|S(0)=S(1)=1]$ in a model with observable confounding as in Model $M_3$ and Assumption \ref{ass_mon1}. For simplicity, we consider the case where selection probabilities, conditional quantile function, and propensity scores are known. For the unknown case and Assumption \ref{ass_mon2}, consider \citeA{semenova2020better} or \citeA{heiler2022hetbounds}.
First note that under Assumption \ref{ass_mon2} $E[Y(1)-Y(0)|S(0)=S(1)=1] = E[Y(1)-Y(0)|S(0)=1]$. We now make use of the conditional independence $Y(d) \perp D|X,S(d)$ derived by $m$-SWIG for model $M_3$ and the counterfactual adjustment formula. 
This yields \begin{align}
    &E[Y(1)|S(0)=1] \notag \\
    &=\sum_x E[Y(1)|S(0)=1,D=1,X=x]P(X=x|S(0)=1) \notag \\
    &=\sum_xE[Y|S(0)=1,D=1,X=x]\frac{P(S=1|D=0,X=x)}{P(S=1|D=0)}P(X=x)  \notag\\
    &\geq \sum_xE[Y|Y \leq q(p_0(x)), S=1,D=1,X=x]\frac{P(S=1|D=0,=x)}{P(S=1|D=0)}P(X=x)  \notag\\
    &=\sum_xE[Y\mathbbm{1}(Y \leq q(p_0(x)))|S=1,D=1,X=x]\frac{P(S=1|D=1,X=x)}{P(S=1|D=0)}P(X=x) 
\end{align}
as $p_0(x) = P(S=1|D=1,X=x)/P(S=1|D=0,X=x)$. This motivates the moment function for the lower bound: 
\begin{align}
    \psi_{L,1}&(W,\eta) = \frac{(\tilde{Y} - E[\tilde{Y}|S=1,D=1,X=x])SD}{P(SD=1|X)}\frac{P(S=1|D=1,X=x)}{P(S=1|D=0)} \notag  \\&+ E[\tilde{Y}|S=1,D=1,X=x]\frac{P(S=1|D=1,X=x)}{P(S=1|D=0)} - E[Y(1)|S(0)=1]
\end{align}
where $\tilde{Y}:= Y\mathbbm{1}(Y \leq q(p_0(x)))$.
For the control outcome, a simple adjustment yields: 
\begin{align}
    E&[Y(0)|S(0)=1] \notag \\ &=\sum_x E[Y(0)|S(0)=1,D=0,X=x]P(X=x|S(0)=1) \notag \\
    &=\sum_x E[Y|S=1,D=0,X=x]\frac{P(S=1|D=0,X=x)}{P(S=1|D=0)}P(X=x) 
\end{align}
which motivates a bias-corrected version of the moment used for conventional \citeA{lee2009training} estimator:
\begin{align}
    \psi_{L,0}&(W,\eta) 
    = \frac{(Y - E[Y|S=1,D=0,X=x])S(1-D)}{P(D=0|X)P(S=1|D=0)} \notag  \\ &+ E[Y|S=1,D=0,X=x]\frac{P(S=1|D=0,X=x)}{P(S=1|D=0)} - E[Y(0)|S(0)=1]
\end{align}
The total effect lower bound can then again be obtained by the difference $\psi_{L,1}(W,\eta) - \psi_{L,0}(W,\eta)$. For the upper bound, steps are analogously with inverted inequalities and complementary trimming shares.

\subsection{Heterogeneous Effects and Bounds} \label{sec_estHET1}
Orthogonal moments can also be used to obtain of heterogeneous effects and effect bounds.
We focus on the case where the heterogeneity variables of interest are (i) known and (ii) low-dimensional but possibly continuous. In particular, we are interested in causal effects or counterfactuals conditional on low-dimensional pre-treatment characteristics $Z$, e.g.~the expected potential outcome or effect of receiving an educational input $d$ at a given level of socio-economic status $Z$ (``group''). We assume that $Z$ are function of observed confounders, i.e.~$Z = f(X)$. This includes subsets of $X$ as a special case.
Consider first the case of a discrete $Z$. A naive approach would to essentially apply the approach from the previous subsection but estimate everything within a given $Z=z$ cell. However, this makes inefficient use of the data if there are similarities across cells in terms of the nuisance functions, e.g.~comparable treatment or selection propensities, or potential outcomes. Thus, we instead suggest to exploit the following identity for counterfactual $E[Y(d)|Z=z]$: \begin{align}
E[Y(d)|Z=z] &=
	E[E[Y(d)|X]|Z=z] \notag \\ &=  E[E[\psi(W,\eta)|X]|Z=z] \notag \\
                        &= E[\psi(W,\eta)|Z=z] \label{eq_ident_hetboundsNP}
\end{align}
and equivalently for quantities other than $E[Y(d)|Z=z]$. 
\eqref{eq_ident_hetboundsNP} implies that we can obtain estimates for \textit{predicted} heterogeneous potential outcomes or causal effects 
from a regression of the respective moment functions $\psi(W,\eta)$ onto $Z$ or variables generated from the dictionary of $Z$.
In the case of discrete $Z$, this boils down to a simple OLS or mean regression using estimate nuisances $\psi(W,\hat{\eta})$ as dependent variable and $Z$ categories as regressors. 
In the point-identified case, confidence intervals can then obtained by standard methods under given dependence/heteroskedasticity assumptions. This approach can also be applied to continuous $Z$ with (potentially misspecificfied) regression or nonparametric/machine learning methods e.g.~series/kernel/forests. For the required high-level conditions consider \citeA{chernozhukov2018generic}, \citeA{semenova2021debiased}, and \citeA{fan2020estimation} or also \citeA{heiler2022hetbounds} for the case of partial identification.  

\subsection{Overlap and Positivity} \label{sec_practicalOverlap1}
Throughout the paper, we have implicitly assumed that conditioning sets are non-empty and any inverse probability weight is different from zero, i.e.~a general form of ``overlap'', also referred to as ``positivity'', holds. Consider Example 1: It requires that $0 < P(D=1|X=x) < 1$ for all $x$ in the support of $X$, i.e.~overlap with regards to the treatment groups \cite{rosenbaum1983central}. For Examples 2 and 3, on the other hand, it is assumed that $P(S=1,D=d|X=x) > 0$ for $d=0,1$ and $x$ in the support of $X$ respectively for identification.

For regular convergence of most estimators and validity of inference, usually a stronger overlap assumption is imposed that bounds these probabilities away from zero by a constant margin \cite{khan2010irregular,HEILER2021valid}. This can empirically be assessed by plotting the distribution of the aforementioned probabilities \cite{HEILER2021valid}. However, such overlap plots and subsequent decisions about trimming and/or the use of robust inference methods should generally be based on the specific probability weights required for the specific parameter of interest, not the generic treatment or selection propensity. We provide an empirical example in Section \ref{sec_SelOverlap1}.

\section{Empirical Analysis} \label{sec_EMPIRICAL1}
In this section we apply some of the methodology to draw inferences from a specific intervention. We evaluate the impact of a teacher’s aide intervention on adolescents' mental health using the Danish teacher's aide experiment \cite{andersen2020effect}. The intervention was a randomized controlled trial in Denmark in 2012/13 using two different types of teacher’s aides to sixth grade classrooms: One treatment used teaching assistants without teaching degrees who spend 14.5 lessons per week in the classroom during nine months. The second treatment used co-teachers with teaching degrees who spend 10.5 lessons per week in the classroom. The costs of the interventions were the same (USD 25,000 per class) because collective agreements required that educated teachers had more preparation time and higher salary.  

\subsection{Teacher's Aides and Mental Health}
\label{coteach_and_mental}
Globally, about 14\% of 10-19-year-olds experience a mental disorder,  which accounts for 13\% of the burden of disease in this group. Depression, anxiety and behavioural disorders are among the leading causes of illness and disability among adolescents.\footnote{\url{https://www.who.int/news-room/fact-sheets/detail/adolescent-mental-health}, accessed June 6, 2023.} Increasing school teachers' time with the individual students may be one way of addressing mental health problems among adolescence. One reason is that with lower student-teacher ratio teachers may be better able to address students' externalizing, in particular disruptive, behavior \cite{mckee2015disruption}. Another reason may be that with lower student-teacher ratio, teachers have more time to observe and adapt teaching to the individual student. This may be particularly important for students with internalizing problems, which may be less visible in the classroom, but highly predictive of anxiety and depression \cite{bryant2020strengths}.

Much of the research on student-teacher ratios focus on class size and most of this research focus on learning outcomes. A couple of studies, though, show positive effects of reducing class size on non-cognitive outcomes such as ``effort, motivation, aspirations, self-confidence, sociability, absenteeism, and anxiety'' \cite{fredriksson2012longterm} and psychological engagement with school (\citeNP{dee2011noncognitive}). However, \citeA{jakobsson2013class} cannot reject no effects on mental health and wellbeing. 

Yet, there are other, more flexible ways of reducing student-teacher ratio than through class size. Especially having an extra teacher in the classroom can be applied in some classrooms some of the time and need not, as reduced class size, be applied in all classes all the time. Furthermore, the effect of student-teacher ratio on mental health is likely dependent on a number of factors such as the quality of the teacher, the types of mental health problems (such as internalizing and externalizing), and characteristics of the students (such as their socioeconomic status). 

There may also be a quantity-quality trade-off. Is more time with lower-educated, lower-paid teaching assistants more effective than less time with a high-educated teacher? 
One way of thinking about the relationship between student-teacher ratio and mental health potentially follows the link between educational production and learning or disruption in class: Following \citeA{lazear2001educational} and \citeA{SCHANZENBACH2020321}, consider a stylized model where $p$ reflects the probability that each child in class behaves, and conversely, $1-p$ the probability that each child misbehaves. Misbehavior may be fighting, talking, or monopolizing the teacher’s time. In a class of $n$ students, $p^n$ is the probability that learning can take place (assuming independence). In this world, teacher’s aides may improve the learning environment through support that increases children’s probability of behaving, $p$, or through reducing effective class size, $n$. \citeA{andersen2020effect} found that teaching assistants were more effective in classrooms with behavioral problems, whereas teacher's aides were more effective when they were used to reduce effective class size, which speaks to such mechanisms.

Furthermore, this quality-quantity trade-off may be conditional on the types of mental health problems among the students. Teachers may be used to and forced into handling externalizing students characterized by hyperactivity and conduct problems. Students' internalizing problems related to emotions and peer relations may be less visible to the teacher and therefore benefit more from a lower student-teacher ratio, which gives the teachers relatively more time with the individual student. Less-educated teaching assistants may be efficient in handling internalizing problems that may require more time compared to externalizing problems that require classroom management skills.

Moreover, the effect of teacher's aides may relate to the parental background of the students. In a human capital investment approach \cite{todd2003specification,fredriksson2016parental}, public and private investments in education and health may be substitutes or complements. For example, high-educated parents may compensate for variation in school inputs meaning that their children are less susceptible to what is going on in school, whereas low-educated parents are less able to do this. \citeA{fredriksson2016parental} find that high-income parents respond to larger class size by spending more time on homework or moving their children to other schools, while low-income parents do not. On the one hand, teacher’s aides clearly reflect a higher level of public investment. If high-educated parents respond by reducing parental investment, the effect of the increased public investment on their children evaporates. \citeA{andersen2020effect} indeed find that students whose parents have no college degree benefit most from teacher’s aides, which is in line with this mechanism. On the other hand, teacher’s aides enable schools to target teaching to individual students; because more resources are available per student and because the doubled teacher inputs may be complementary in speaking to different individual students or different needs. All these mechanism have the potential to raise basic skills and improve the learning environment which can be related to mental health as well. \citeA{andersen2020effect} show that teaching assistants were more effective in classrooms with behavioral problems and interpret this finding as indicating that complementarities explain why teaching assistants were effective at improving reading skills. 

Finally, mechanisms for mental health may differ compared to educational production and between externalizing and internalizing domains in particular due to (in)attention of parents and teachers towards internalizing problems. \citeA{novak2018prevention} argue that teachers and even parents often do not recognize internalizing problems in exact amount. Children and youth tend to hide them and avoid asking for help. Schools and teachers thus can play an important role in reducing internalizing problems, including promotion of emotional well-being and first symptom response.


We exploit the Danish teacher's aide experiment to address these questions about the conditional effects of reducing student-teacher ratio on student mental health. In a previous analysis of the experiment, \citeA{andersen2020effect} study the impact of the two types of teacher’s aides on test scores immediately after the intervention and two years later. They find that both types of aides have positive impacts on test scores and the effects are persistent over time for students of parents without a college degree. Main results show no significant difference between the two treatment arms (Table 4 in \citeA{andersen2020effect}). Suggestive evidence showed that the teaching assistants were most effective in classrooms with behavioral problems and when they coordinated tasks with the main teacher, whereas the teacher's aides were most effective when they were used as a means to flexible class-size reduction.

\citeA{Andersenetal2022} also investigate whether teacher’s aides increased inclusion of students with special education needs (SEN) in regular classrooms and whether they influenced the academic achievement of SEN students throughout compulsory education. Teacher’s aides resulted in 6--7 percentage points higher inclusion, and treated SEN students were able to stay in regular classrooms throughout compulsory education. Academic gains were present in the short run but disappeared over time. 

Similar teacher's aide experiments were also carried out in Project STAR Tennessee in 1985/89 \cite{schanzenbach2006have} and in Norway 2013/14 \cite{borgen2022funds} and in 2016/17–2019/20 \cite{bonnesronning2022smallgroup}. For Project STAR, positive effects of class-size reduction on academic outcomes are reported but no effects of teacher’s aides (who were mostly used for practical support, though). For Norway, \citeA{borgen2022funds} find no effects of extra resources for teachers on academic achievement, but some favorable effects on behavioral and other outcomes unrelated to academic achievement. \citeA{bonnesronning2022smallgroup} find positive effects of small-group instruction on math skills.

\subsection{Experiment and Data}
\subsubsection{Experiment}
\label{sec_exper}

Conditional on predicted school average test scores, we divided 105 participating schools in 35 strata. We randomly allocated one school per stratum to each of the two treatments (i.e. teacher's aides with or without a teaching degree) and one school per stratum to a control group. 
The intervention was in place for roughly 85\% of the school year, from October 1, 2012 to June 20, 2013. The intervention and control groups were announced on August 15, 2012, leaving the schools 1.5 months to search for and employ the respective teacher’s aides. 

The requirements in terms of how to make effective use of the teacher’s aides were extremely flexible, allowing the school principal to take into account the particular group of students in each cohort. The teacher’s aides mainly supported achievement or wellbeing and did not supply much practical support \cite{andersen2020effect}. About 80\% of the teachers answer that the primary task of the teacher’s aides was to improve student achievement or wellbeing (avoiding conflict and the like) with student achievement
being more dominant for co-teachers with a teaching degree than for teaching assistant
without a degree.\footnote{A total of 75\% of the main teachers reported that the primary task of the co-teacher w/degree was student achievement and 6\% reported wellbeing, whereas the comparable numbers for the teaching assistant w/o degree were 51\% and 29\%, respectively. See \citeA{andersen2020effect}, Table 1.} Three quarters of the aides across the two types report spending time on handling conflicts and other disruptions among students. Conflict, peer relations, and disruptive behavior are strong predictors of adolescent mental health \cite{goodman2000using,kristoffersen2015disruptive}. 
For this particular study, we analyse both treatment arms as well as an aggregated single teacher's aide treatment that combines both treatment arms.

\subsubsection{Data}
We use the sample from the original trial evaluation in \citeA{andersen2020effect}. This includes 5213 students from 105 schools registered as participants at the start of the trial. All participants are registered with unique tracking IDs and surveyed in the beginning and end of the intervention period to collect student-detailed information on outcome and background measures.


To construct outcome measures of child mental health, we use the strength and difficulties questionnaire (SDQ), which reflects emotions, behaviors and peer relations of children and young people. 
The SDQ questionnaire consists of 25 questions and can be used to capture the perspective of parents, teachers or youth aged 11 years and above (see \citeA{goodman1997strengths}, \citeA{goodman2000using}, and \url{sdqinfo.org}).  We use the self-reported version of SDQ as students are above 11 years old. The questions are divided between five subscales, each consisting of five items: 1) hyperactivity, 2) conduct problems, 3) emotional symptoms, 4) peer problems, and 5) pro-social scale. 
The score in each subscale ranges from 0 and 10. 
 We use the broader scales for externalizing and internalizing behavior constructed by the sum of subscales 1-2 and 3-4 for externalizing and internalizing behavior, respectively, ranging from 0 to 20. The range 0 to 4 to 5 is considered normal in Danish children age 10-12 (see \url{sdqinfo.org}).\footnote{\citeA{goodman2010use} investigate when to use the five detailed subscales versus the combined scales for externalizing and internalizing behavior. They find that it is difficult to distinguish hyperactivity from conduct problems in low-risk populations, and similarly, difficult to distinguish emotional symptoms from peer problems in low-risk populations, and therefore the broader scales for externalizing and internalizing behavior are more appropriate for such populations. However, the detailed scales may be retained when screening for psychiatric disorders in high-risk populations.} We standardize the scales; one standard deviation corresponds to 3.2 and 3.5 points for internalizing and externalizing measure, respectively.

We link survey data from students with register data from Statistics Denmark on students and their parents. This provides us an extended set of covariates that are pre-determined and less subject to missingness concerns as they are from administrative registers, an advantage we leverage to inform the selection model.

For our heterogeneity analysis, we focus on student socio-economic status (SES) as measured by parents' employment status and college education as in \citeA{andersen2020effect}. We distinguish whether both, one, or none of the parents are employed and whether at least one parent has a college degree as opposed to none. Other employment or education information is absorbed by their last, i.e.~lowest, categories.

\begin{table}[!h]
    \centering
    \begin{threeparttable}
    \caption{SDQ Survey Response Rates}
    \label{tab_responserates} \footnotesize
    \begin{tabular}{lccccr}
          \hline \hline
          \\[-1ex]
    & Total & Control  				& Co-teacher 						& Teaching Assistant \\
	  &     &  group				& w/degree					& w/o degree 	 \\ \hline \\[-1ex]
 Full Sample&       85.2\% & 78.0\% & 89.5\% & 88.4\%  \\ \\[-.5ex]
 Employment & \\ \hline \\[-.5ex]
both & 87.0\% &  80.5\% &  89.7\% &  91.1\%  \\
one & 82.7\% &  76.2\% &  85.6\% &  86.4\% \\
none/missing & 77.7\% &  65.5\% &  84.6\% &  85.1\% \\ \\[-.5ex]
 College Education & \\ \hline \\[-.5ex]
yes & 86.4\% &  76.8\% &  90.8\% &  91.6\% \\
none/missing & 84.4\% &  78.9\% &  86.5\% &  88.1\% \\
 \\[-1ex] \hline \hline
    \end{tabular} \begin{justify} \footnotesize
    Response rates across groups of the post-intervention SDQ survey conducted June 2013 for the full sample and stratified by socio-economic status variables parental employment and college education.         
    \end{justify}
    \end{threeparttable}
\end{table}

Table \ref{tab_responserates} contains the different response rates for the outcome measures from the post-intervention survey across groups. 
The measurements are subject to significant, differential attrition rates. Thus, ignoring potentially systematic non-response could bias results. There are multiple reasons for missing outcome data in this context. First, the student is absent on the day of survey collection. Second, the student has moved to another school that is not part of the trial. If the student moves to another school part of the trial, the student will participate in survey collection at this school. In the evaluation, we use students’ initial treatment assignment. Both reasons for missingness seem to be at least in part predictable. In particular, the response rates are highly heterogeneous across SES, in particular with regards to employment status of the parents. For employment status, the largest differences in attrition rates within groups range from 5.1\% to 15\%.  We will exploit these differences in the heterogeneity analysis to make more precise statements about subgroup specific effects under weak assumptions. 

Tables \ref{tab_app_desc_11}--\ref{tab_app_desc_22} in Appendix \ref{app_empirical} report all the different sample characteristics.\footnote{We use an extended set of covariates covering detailed information on students' birth information, demographics, and school performance as well as parental education, year of work experience, employment status, and health. All this information is collected before treatment allocation. All non-outcome data is from Danish administrative data.} Students' average internalizing and externalizing SDQ scores are 4.39 and 5.03, respectively. Considering the subgroup indicators used for the heterogeneity analyses, 42 percent of the sample have at least one college-educated parent and 70 percent have both parents employed. 
In general, the characteristics are representative of students in Danish public schools. Previous research documents that the randomization was successful and that the sample was well balanced across treatment arms, with the exception of increased inclusion of SEN students in
treatment classrooms \cite{Andersenetal2022}. 

Using these covariates, we also test for determinants of attrition and differential attrition as discussed in Section \ref{sec_testattrition1}. Both tests indicate significant determinants and differences ($p < 0.01$). This is a first indication of non-random attrition. However, as these tests are only indicative and neither necessary nor sufficient for rejecting no selection bias, as discussed in Section \ref{sec_testattrition1}, we apply a selection of models in what follows and compare results under increasingly weak assumptions. 

\subsubsection{Models and Estimation} \label{sec_modelsandesti1}
We estimate the effects of the teacher's aide intervention on SDQ internalizing and externalizing scores using various models. Table \ref{tab:models} contains an overview of all models and what type of selection process they allow for in the outcome process in reference to Section \ref{sec_methodology1_selandmiss}. Model $M_1$ yields a standard naive mean comparison. Model $M_2(D)$ is equal to Model $M_2$ but without including confounders $X$, i.e.~only treatment based missingness is taken into account. Models $M_2$, $M_3$, and $M_3(ZRLee)$ are discussed in Section \ref{sec_methodology1_selandmiss}.

\begin{table}[!h]
    \centering 
    \begin{threeparttable}
    \caption{Models and Assumptions}
    \label{tab:models} \footnotesize
    \begin{tabular}{lcccccc}
    \hline \hline \\[-1ex]
     \centering Model & \multicolumn{3}{c}{Selection on} & Missingness & Uses $X$ & Monotonicty \\
         & Treatment & Observables & Unobservables   &  &  &    \\ \hline \\[-1ex] 
      $M_1$ & & & &  MCAR &\\
      $M_2(D)$ & \checkmark & & &  MAR($D$) &\\
      $M_2$ & \checkmark & \checkmark&  & MAR($X,D$) & \checkmark & \\
      $M_3(ZRLee)$ & \checkmark & \checkmark&  \checkmark & MNAR  & & Assumption~\ref{ass_mon1}\\
      $M_3$ & \checkmark & \checkmark&  \checkmark & MNAR &\checkmark &Assumption~\ref{ass_mon2}  \\ \hline \hline
    \end{tabular}
          \begin{justify}
          This table contains the models used for estimation, underlying exclusion restrictions, and missingness classifications. For the missing-at-random cases, denote $D$ = Treatment, $X$ = Observables = (Stratum, Covariates). Stratum here describe the variable used for stratification before randomization of the treatment (predicted school performance in 6th grade). ZRLee refers to the Zhang-Rubin-Lee bounds without covariates \cite{zhang2003estimation,lee2009training}.
      \end{justify}
      \end{threeparttable}
\end{table}

For the models that require nuisance parameter estimation, we use orthogonal moment functions and cross-fitting 
related to Section \ref{sec_estimation} but treating all nuisances as unknown.\footnote{We use the score functions that account for estimation of the nuisance functions as in \citeA{chernozhukov2018double} and \citeA{heiler2022hetbounds} for $M_2$ and $M_3$.}  The latter are estimated via honest generalized random forests ($M_2$: probability forests and regression forests, $M_3$: like $M_2$ + quantile forests) with default tuning parameters from package \verb|grf|.\footnote{We evaluated the tuning parameters choices in terms of typical validation criteria and found only little room for optimization when changing the number of predictors for subsampling. Thus we resort to their default values. Estimates using restricted parametric models are qualitatively similar and available upon request.} We employ 10-fold cross-fitting. As this is a randomized experiment, propensity scores are known and constant. For models $M_1$, $M_2(D)$, and $M_2$, we provide standard heteroskedasticity robust 95\% confidence intervals, while for $M_3$ we use improved misspecification and heteroskedasticity robust confidence intervals at a 90\% level as this method is conservative \cite{heiler2022hetbounds}. We provide the effects for both treatment arms separately as well as the aggregate treatment.

Differences in results beyond sampling variation can occur due to varying missingness assumption, (unaccounted) heterogeneity, or misspecification. 
We deem monotonicity, in particular Assumption \ref{ass_mon2}, as credible in this application as the propensity to attend school under the (non)presence of a teacher's aide is either explained by systematic socio-demographics related to attendance and performance in school (observed) or attendance preference in relation to the particular teacher (or induced change of school). Monotonicity then says we can reliably predict the sign of the latter from the administrative register information. Given the large set of register information, we also do not completely rule out the likelihood of MAR/selection on observables. Thus, our ex-ante preference relation over models is $M_3 \succ M_3(ZRLee) \succ M_2(X,D) \succ M_2(D) \succ M_1$.

\subsubsection{Selection Overlap} 
\label{sec_SelOverlap1}

\begin{figure}[!h]
    \centering
    \caption{Selection Probabilities: Kernel Density Estimates}
    \label{fig_kernel_sds1}
    \includegraphics[width=0.8333333\textwidth, trim = 0 120 0 120, clip]{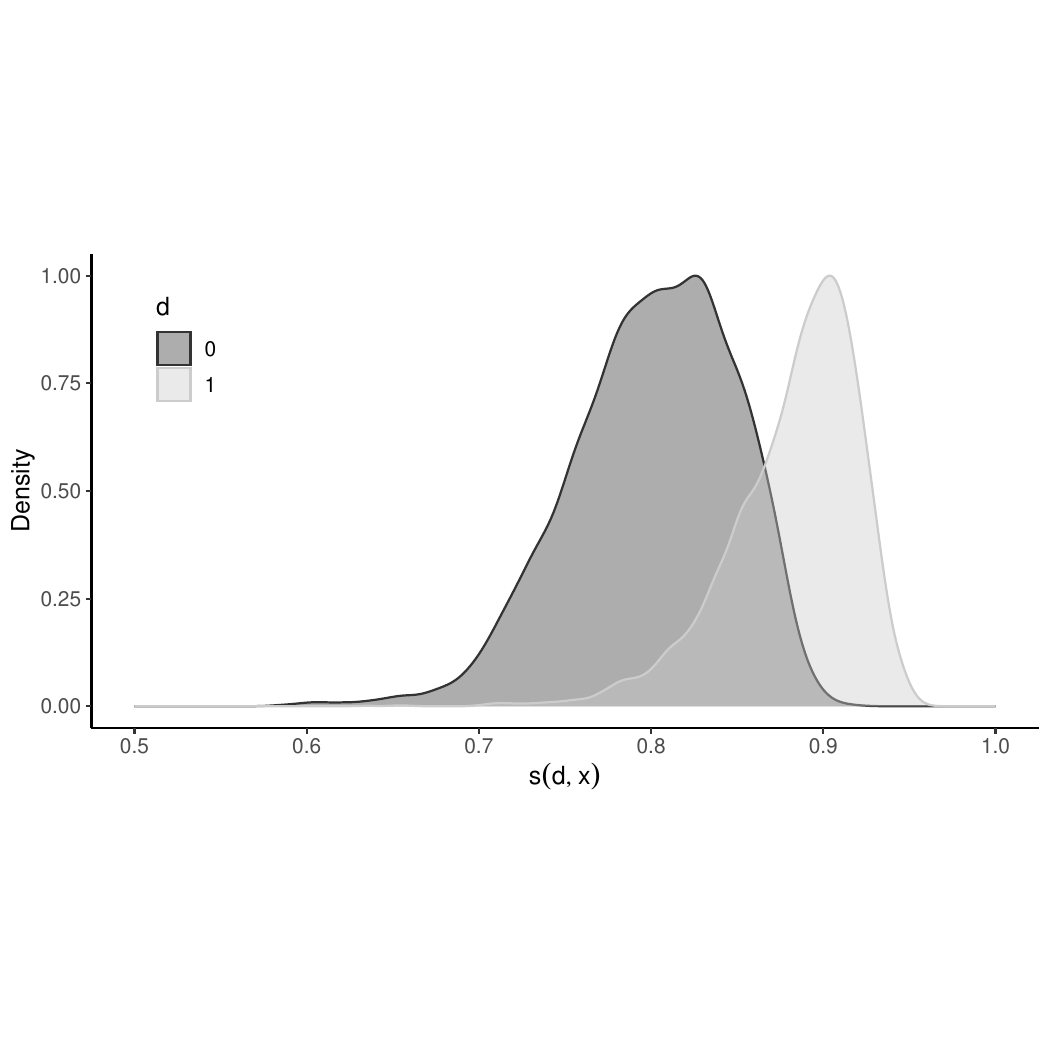}
    \begin{justify} \footnotesize
        This figure contains the normalized kernel density estimates for the predicted selection probabilities conditional on treatment status. Estimates are obtained using probability forests and 10-fold cross-fitting. 
    \end{justify}
\end{figure}

Overlap with regards to the treatment was assured by the randomization protocol \cite{andersen2020effect}. Given the potentially non-random selection, however, this might not be enough for identification. 
In particular, models $M_2$ and $M_3$ require comparable units for selection and treatment jointly, i.e.~overlap $P(S=1,D=d|X=x)>0$ for $d =0,1$ at any $x$, see Section \ref{sec_practicalOverlap1}. Due to treatment randomization $P(D=d|X=x) = P(D=d)$. Thus, joint overlap is equivalent to $P(S=1|X=x,D=d) > 0$ for $d =0,1$ and any $x$.

Figure \ref{fig_kernel_sds1} depicts the kernel density estimates of the estimated selection probabilities given $d=0,1$. The probabilities concentrate in the areas between 0.6 and 0.95 and are all well separated from the boundary. Thus, no trimming or other form of correction is required for regular inference in what follows \cite{HEILER2021valid}. 
Overall, the treatment shifts the distribution to the right. For most students, the effect of treatment on selection is positive. The estimated propensity ratios show that 98.77\% are predicted to be of the positive monotonicity type with the other negative type of 1.23\% (64 students) belonging to a particular subset of schools/teachers. Thus, even the unconditional monotonicity Assumption \ref{ass_mon1} could serve as a good approximation for the majority of the sample.

\afterpage{
    \begin{landscape}
\begin{table}[t] 
\centering 
\begin{threeparttable}
\caption{Internalizing and Externalizing Scores: Unconditional Estimation Results}  \label{tab:internal_external}
\scriptsize
\begin{tabular}{cc}
\begin{tabular}{lccc} \hline \hline
\\[-0.5 ex] 
& \multicolumn{3}{c}{Internalizing Scores} \\\cline{2-4} \\[-0.5ex]
    & Aggregate  				& Co-teacher 						& Teaching Assistant 	\\
	&  				& w/degree					& w/o degree 	 \\ \hline \\[-1ex]
	$M_1$ 	& -0.0703 				& -0.0515 					& -0.0917 \\
	& (-0.1337,~-0.0068)	& (-0.1235,~0.0204)			& (-0.1651,~-0.0182)	\\ \\[-1ex]
	$M_2(D)$	& -0.0714				& -0.0512					&-0.0919\\
	& (-0.1358,~-0.0069)	& (-0.1231,~0.0207)			& (-0.1656,~-0.0183)\\ \\[-1ex]
	$M_2$	& -0.0671				& -0.0522					&-0.0834\\
	& (-0.1296,~-0.0046)	& (-0.1227,~0.0182)			& (-0.1551,~-0.0118)\\ \\[-1ex]
	$M_3(ZRLee)$ & [-0.3155,~0.0589]		& [-0.3001,~0.0746]			&[-0.3330,~0.0409]\\
	&	(-0.3490,~0.0911)& (-0.3405,~0.1161)			&(-0.3756,~0.0833) \\ \\[-1ex]
	$M_3$ & [-0.2388,~0.0114]			& [-0.4013,~0.1409]			&[-0.0947,~-0.0871]\\
	&(-0.3325,~0.0778)&(-0.5195,~0.2161) &(-0.1739,~0.0481)			\\[1ex] \hline \\[-1ex] 
 $n_{treat}$ & 3424 & 1814 & 1610 \\
 $n_{control}$ & 1789 & 1789 & 1789 \\[1ex]
 \hline \hline 
\end{tabular} 

& \begin{tabular}{lccc} \hline \hline
\\[-0.5 ex] 
& \multicolumn{3}{c}{Externalizing Scores} \\\cline{2-4} \\[-0.5ex]  
& Aggregate  				& Co-teacher 						& Teaching Assistant 	\\
	&  				& w/degree					& w/o degree 	 \\ \hline \\[-1ex]
	$M_1$ 	& -0.0290 				& -0.0077 					& -0.0533 \\
	& (-0.0928,~0.0347)	& (-0.0797,~0.0642)			& (-0.1271,~ 0.0206)	\\ \\[-1ex]
	$M_2(D)$	& -0.0295				& -0.0075					&-0.0526\\
	& (-0.0942,~0.0353)	& (-0.0795,~0.0644)			& (-0.1266,~0.0215)\\ \\[-1ex]
	$M_2$	& -0.0160				& 0.0032					&-0.0357\\
	& (-0.0775,~0.0455)	& (-0.0657,~0.0721)			& (-0.1064,~0.0349)\\ \\[-1ex]
	$M_3(ZRLee)$ & [-0.2760,~0.1001]		& [-0.2587,~0.1150]			&[-0.2957,~0.0829]\\
	&	(-0.3096,~0.1341)& (-0.2993,~0.1581)			&(-0.3387,~0.1271) \\ \\[-1ex]
	$M_3$ & [-0.1964,~0.0725]			& [-0.3342,~0.1942]			&[-0.0659,~-0.0360]\\
	&(-0.2898,~0.1406)&(-0.4529,~0.2709) &(-0.1998,~0.0452)			\\[1ex] \hline \\[-1ex] 
 $n_{treat}$ & 3424 & 1814 & 1610 \\
 $n_{control}$ & 1789 & 1789 & 1789 \\[1ex] \hline \hline 
\end{tabular} 
\end{tabular}
\begin{justify} \scriptsize
	Point and interval estimates of the causal effect on internalizing and externalizing scores for aggregate treatment and each treatment arm for all models. Regular brackets are confidence intervals, interval brackets are estimated identified sets. All results are estimated in standard deviations over the observed study population.  
\end{justify}
\end{threeparttable}
\end{table}
        \begin{table}[!h]
    \centering 
    \begin{threeparttable}
    \caption{Internalizing Scores for Teaching Assistant: Heterogeneity Analysis by SES}
    \label{tab:SEShetero_int1} \tiny
    \begin{tabular}{lccccS[table-format=3.2]cS[table-format=3.2]cS[table-format=3.2]cr} \hline \hline \\[-0.5ex]
    & \multicolumn{2}{c}{$M_3$}  & \multicolumn{2}{c}{$M_3(ZRLee)$} & \multicolumn{2}{c}{$M_2$}  & \multicolumn{2}{c}{$M_2(D)$} & \multicolumn{2}{c}{$M_1$}   &  \\ 
         & Estimate & CI &  Estimate & CI  & \text{Estimate} & CI &  \text{Estimate} & CI & \text{Estimate} & CI &  $n$  \\ \hline \\[-1ex]
    Full sample & [-0.0947,~-0.0871] & (-0.1739,~0.0481) & [-0.2587,~0.1150] & (-0.3387,~0.1271) & -0.0357	& (-0.1064,~0.0349)		&			-0.0526	& (-0.1266,~0.0215)		&			-0.0533	& (-0.1271,~ 0.0206)		&
 3399 \\ \\[-1ex]
    Employment & &\\ \hline \\[-1ex]
    both & [-0.1353,~-0.1167] & (-0.2840,~-0.0421) & [-0.2833,~0.0603] &   (-0.3288,~0.1071) & -0.0296	&(-0.1933,~0.1341)&	-0.1391	&(-0.3094,~0.0312)&	-0.0697	&(-0.1528,~0.0134)&
2524 \\
    one & [0.0696,~0.1300]  & (-0.2242,~0.3173) & [-0.1657,~0.1838] &  (-0.2703,~0.2840) & -0.0458	&(-0.3719,~0.2803)&	0.3440	&(0.0202,~0.6678)&	0.0650	&(-0.1016,~0.2316)&
704 \\
    none/m & [-0.2533,~-0.1883] & (-0.5199,~-0.0072) &  [-0.6364,~0.0692] &  (-0.7959,~0.2133) & 0.3553	&(-0.1370,~0.8476)&	1.1419	&(0.6821,~1.6017)&	-0.1673	&(-0.4209,~0.0863)&
375 \\ \\[-1ex]
    College & &\\ \hline \\[-1ex]
    yes  &[-0.3861,~0.0308] & (-0.5735,~0.1445) & [-0.3556,~0.0750] &   (-0.4114,~0.1393) & -0.1778	&(-0.3918,~0.0362)&	-0.3696	&(-0.5877,~-0.1515)&	-0.0781	&(-0.1869,~0.0307)&
1467 \\
    no/m   &[-0.1584,~0.0171] & (-0.1853,~0.1288) & [-0.2328,~0.0760] &  (-0.2912,~0.1302) & 0.1371	&(-0.0495,~0.3237)&	0.4053	&(0.2156,~0.5950)&	-0.0274	&(-0.1223,~0.0675)&
2136 \\ \\[-1ex]
    \hline \hline 
    \end{tabular} 
    \begin{justify} \scriptsize
	Unconditional and heterogeneous interval estimates of the causal effect on internalizing scores for teaching assistant treatment using all models. Regular brackets are confidence intervals, interval brackets are estimated identified sets. All results are reported in standard deviations over the observed study population.
\end{justify}
\end{threeparttable}
\end{table} 

    \end{landscape}
}

\subsection{Results}
\subsubsection{Unconditional Effects}

Table \ref{tab:internal_external} contains effect and effect bound estimates for internalizing scores (left panel). For the aggregate treatment, models $M_1$, $M_2(D)$, and $M_2$ provide a treatment effect of \mbox{-0.07} SD suggesting that teacher's aides reduce students' internalizing difficulties. When weak assumptions are imposed in $M_3(ZRLee)$, bounds are relatively wide and include sizeable negative effects as moderate effects above zero. With additional covariate information and weakest assumptions $M_3$, the bounds tighten to $[-0.2388, 0.011]$.

There are relevant differences between the treatment arms. The point-identified approaches suggest an effect of -0.05 SD for the co-teacher intervention with degree and around -0.09 SD for the teaching assistant without degree.  
Moreover, while the bounds using both $M_3$ specifications are rather wide for the co-teacher w/degree arm, this is not the case for the teaching assistant w/o degree arm. Here the bounds under the weakest assumptions, $M_3$, tighten to  $[-0.095,-0.087]$ close to the other point-identified models that do not control for unobservables.


Table \ref{tab:internal_external} also shows the results for externalizing scores (right panel). Qualitatively, point and interval estimates are very consistent with those for internalizing scores, albeit of smaller magnitude. In particular, the point estimates are about 0.04 SD smaller for all specifications. Moreover, all confidence intervals now include zero indicating that we cannot conclude with precision that the intervention significantly reduced externalizing behavior on average despite all point estimates and the $M_3$ interval estimates being strictly negative. This could be due to a lack of power in detecting these smaller effects compared to internalizing scores.

\subsubsection{Conditional Effects}
We further analyze the effect bounds 
as a function of socio-economic status of the student's parents (employment and education). 
In particular, we estimate heterogeneous effects and effect bounds using all models with the approach outlined in Section \ref{sec_estHET1}. The focus is on teaching assistant without degree and internalizing scores due to the stronger signal in the unconditional analysis. 

Table \ref{tab:SEShetero_int1} contains point and set estimates and confidence intervals for the teaching assistant effect on internalizing scores using the two SES measures and their respective sample sizes. 
Our preferred specification $M_3$ suggests that there is a clear signal using the employment measure:  In particular, bounds are narrow and significant with interval estimates of -0.135 to -0.117 SD for children with both parents employed and -0.253 to -0.188 SD for children without employed parents respectively. For the latter, the confidence interval is very close to zero. Given the small sample size in this group ($n=375$), we interpret this result with caution. The medium category (one) has a positive estimated interval, but is statistically insignificant. 
The relatively narrow width of the interval for the the fully employed group in $M_3$ can be explained by the lower frequency of missing outcome data. This subgroup has the lowest missingness rate in the control and the second-lowest in the treatment group, see Table \ref{tab_responserates}. 

Using the college measure, estimates are generally statistically insignificant. 
The estimates obtained from $M_1$ and $M_3(ZRLee)$ are consistent with the intervals from $M_3$. However, likely due to the lack of covariate information, they are statistically insignificant. The missing at random models $M_2$ and $M_2(D)$, on the other hand, show clear signs of misspecification. In particular, $M_2(D)$ provides unreasonably large effects for both SES measures in several subgroups. For instance, it suggest a 1.142 SD increase in the none/m category for employment which can be rejected when comparing it to the confidence intervals obtained under weaker assumptions.  


Such misspecification would be masked when only considering unconditional estimates. Thus, the analysis highlights the necessity and utility of careful specification search, rich covariate information, and relaxation of missingness assumptions.

\subsection{Discussion of Results}
Mental health issues among adolescents is an increasing problem in many modern societies. Thus, studying the potential positive and negative effects of traditionally human capital oriented interventions is of increasing importance. Measures of mental health, well-being, and behavior in particular outside of a clinical context are regularly plagued by the data issues addressed in this study. In this section, we discuss the results of the empirical analyses of the impact of teacher's aides on adolescent mental health. Our results point towards three findings. 

First, effects of teacher's aides on internalizing behavior are numerically higher and tend to be more precisely estimated compared to effects on externalizing behavior. The consistent differences is indicative of genuine effect differences and not just an artifact of differences in related socio-economic status, pre-intervention health, academic performance, or else. Teacher's aides represent a flexible way of lowering the student-teacher ratio, which may leave teachers more time to recognize and accommodate students with internalizing problems, as discussed in Section \ref{coteach_and_mental}. While a lower student-teacher ratio should also allow teachers to better cope with externalizing behavior in students, such a conclusion is not supported with high precision. Externalizing problems could also require differently targeted or specialized interventions.

Second, teaching assistants without a degree tend to be more effective in improving mental health than co-teachers with a degree. At a constant dose-response relationship, about half of the total difference between co-teacher and teaching assistant effect could be explained by the differences in treatment intensity (10.5 vs 14.5 hours/week). In addition to high dosage, the teaching assistant treatment is characterized by more frequently having the primary focus on wellbeing (29\% versus 6\%, see subsection \ref{sec_exper}), which may explain its relative effectiveness in reducing internalizing behavior compared to the co-teacher w/degree.

Third, the results suggest that there are benefits of teaching assistants on internalizing behavior for a large population of students. In particular, we detect benefits for students with an economically more advantaged background. While our own previous research on the impact of the teacher's aides interventions show that primarily more disadvantaged students benefit in terms of persistently higher reading scores \cite{andersen2020effect}, the current results suggest that also more advantaged students tend to benefit when it comes to improved mental health. The estimates of effects for students with only one employed parent are closer to zero and more imprecise, which may either reflect smaller genuine effects or more missing data in this group. The precise mechanism requires more research.  



\section{Concluding remarks} \label{sec_CONCL1}
Surveys on mental health and well-being are only one example where missing data problems pose a threat to the credibility of evaluations studies even when collected in the context of a randomized control trial. 
The presented methodology serves as a general framework for handling such and more general evaluation problems with missing data and is accessible for researchers with various methodological backgrounds. There are many other possible avenues for a fruitful synthesis of counterfactual, graphical, and classic econometric methods in the realm of causal inference outside of the problems considered in this paper. 

\newpage

\bibliographystyle{apacite}
\renewcommand{\APACrefYearMonthDay}[3]{\APACrefYear{#1}}
\bibliography{bounds2,bounds1,coTeach1}

\appendix
\section{Supplementary Material to Section \ref{sec_estimation}} \label{sec_App_moments1}
\subsection{Derivations of Moment Functions}
The moment functions follows by simple application of the law of iterated expectations. In general, we have that, \begin{align*}
    E[YS\mathbbm{1}(D=d)|X] = E[Y|S=1,D=d,X]P(S=1,D=d|X)
\end{align*}
We use this formula throughout to add and subtract bias corrections.
\subsubsection{Example 1: ATE without Selection}
Here we always have $S=1$. The bias correction is thus obtained by noting that \begin{align*}
    E\bigg[\frac{Y\mathbbm{1}(D=d)}{P(D=d|X)}\bigg|X\bigg] 
    &= E[Y|D=d,X]\frac{P(D=d|X)}{P(D=d|X)} \\ 
    &= E[Y|D=d,X]
\end{align*}
using the law of iterated expectations and adding the corresponding term to the simple counterfactual adjustment formula.

\subsubsection{Example 2: ATT in Model $M_2$}
Here we have selection and conditioning on the treated $D=1$. However, analogously to Example 1,
\begin{align*}
E\bigg[YSD&\frac{P(D=1|X)}{P(SD=1|X)P(D=1)}\bigg|X\bigg] \\
&=E[Y|S=1,D=1,X]P(SD=1|X)\frac{P(D=1|X)}{P(SD=1|X)P(D=1)} \\
&=E[Y|S=1,D=1,X]\frac{P(D=1|X)}{P(D=1)}
\end{align*}
which then cancels with the weighted adjustment term in the moment function in expectation. For the control part the derivation is equivalent but with modified weights.

\subsubsection{Example 3: Zhang-Rubin-Lee Bounds in Model $M_3$}
Even though there is trimming in $\tilde{Y} = Y\mathbbm{1}(Y \leq q(p_0(X)))$, the bias correction follows the same logic as in Example 2 but with different weights.
\begin{align*}
    E\bigg[\tilde{Y}SD&\frac{P(S=1|D=1,X)}{P(SD=1|X)P(S=1|D=0)}\bigg|X\bigg] \\
    &= E[\tilde{Y}|S=1,D=1,X]P(SD=1|X)\frac{P(S=1|D=1,X)}{P(SD=1|X)P(S=1|D=0)} \\
    &= E[\tilde{Y}|S=1,D=1,X]\frac{P(S=1|D=1,X)}{P(S=1|D=0)}
\end{align*}
The moment function follows from adding and subtracting the corresponding term inside the brackets and its conditional expectation respectively. 

\section{Supplementary Material to Section \ref{sec_EMPIRICAL1}} \label{app_empirical}
\begin{table}[!ht] 
    \centering \caption{Descriptive Statistics for Full and Control Samples:  Part 1}\label{tab_app_desc_11} \scriptsize
    \begin{tabular}{lllllll}
    \hline \hline
        ~ & \multicolumn{3}{c}{All students} &  \multicolumn{3}{c}{Control group}  \\ 
        ~ & $n$ & mean & sd & $n$ & mean & sd \\ \hline \\[-0.5ex]
        \textbf{Student outcomes} & ~ & ~ & ~ & ~ & ~ & ~ \\ 
        SDQ internalizing score & 4443 & 4.39 & 3.23 & 1396 & 4.55 & 3.24 \\ 
        SDQ internalizing score, missing  & 5213 & 0.148 & ~ & 1789 & 0.220 & ~ \\ 
        SDQ externalizing score & 4443 & 5.03 & 3.50 & 1396 & 5.10 & 3.54 \\ 
        SDQ externalizing score, missing & 5213 & 0.148 & ~ & 1789 & 0.220 & ~ \\ 
        ~ & ~ & ~ & ~ & ~ & ~ & ~ \\ 
        \textbf{Subgroup indicators} & ~ & ~ & ~ & ~ & ~ & ~ \\ 
        SES education: Yes & 5213 & 0.42 & ~ & 1789 & 0.41 & ~ \\ 
        SES education: No/m & 5213 & 0.58 & ~ & 1789 & 0.59 & ~ \\ 
        SES employment: Both & 5213 & 0.70 & ~ & 1789 & 0.70 & ~ \\ 
        SES employment: One & 5213 & 0.20 & ~ & 1789 & 0.19 & ~ \\ 
        SES employment: None/m & 5213 & 0.10 & ~ & 1789 & 0.11 & ~ \\ 
        ~ & ~ & ~ & ~ & ~ & ~ & ~ \\ 
        \textbf{Student birth information} & ~ & ~ & ~ & ~ & ~ & ~ \\ 
        Boy & 5213 & 0.51 & ~ & 1789 & 0.51 & ~ \\ 
        Birthweight (grams) & 5213 & 3274 & 1068 & 1789 & 3261 & 1075 \\ 
        Birthweight low (0/1) & 5213 & 0.06 & ~ & 1789 & 0.06 & ~ \\ 
        Birthweight high  (0/1) & 5213 & 0.04 & ~ & 1789 & 0.03 & ~ \\ 
        Premature week 32 (0/1) & 5213 & 0.01 & ~ & 1789 & 0.01 & ~ \\ 
        Mother's age at birth & 5213 & 28.66 & 6.43 & 1789 & 28.60 & 6.54 \\ 
        Father's age at birth & 5213 & 30.42 & 8.97 & 1789 & 30.43 & 9.05 \\ 
        Birth order by mother & 5213 & 1.81 & 1.00 & 1789 & 1.79 & 0.95 \\ 
        Birth order by father & 5213 & 1.81 & 1.05 & 1789 & 1.79 & 1.03 \\ 
        Twin, triplet etc. & 5213 & 0.03 & ~ & 1789 & 0.04 & ~ \\ 
        Student demographics & ~ & ~ & ~ & ~ & ~ & ~ \\ 
        Origin: Danish & 5213 & 0.87 & ~ & 1789 & 0.85 & ~ \\ 
        Origin: Nonwestern & 5213 & 0.11 & ~ & 1789 & 0.13 & ~ \\ 
        Origin: Western & 5213 & 0.01 & ~ & 1789 & 0.01 & ~ \\ 
        Origin: missing & 5213 & 0.01 & ~ & 1789 & 0.01 & ~ \\ 
        Living with both parents & 5213 & 0.65 & ~ & 1789 & 0.64 & ~ \\ 
        Living with single mother & 5213 & 0.24 & ~ & 1789 & 0.24 & ~ \\ 
        Placed out of home & 5213 & 0.01 & ~ & 1789 & 0.01 & ~ \\ 
        \\
        \textbf{Student school information} & ~ & ~ & ~ & ~ & ~ & ~ \\ 
        Reading score from grade 4 & 4967 & 0.03 & 0.97 & 1690 & -0.01 & 0.96 \\ 
        Reading score, missing & 5213 & 0.05 & ~ & 1789 & 0.06 & ~ \\ 
        Math score from grade 3 & 4494 & 0.02 & 1.00 & 1513 & -0.01 & 0.98 \\ 
        Math score, missing & 5213 & 0.14 & ~ & 1789 & 0.15 & ~ \\ 
        Special education needs & 5213 & 0.13 & ~ & 1789 & 0.14 & ~ \\ 
        Special education class & 5213 & 0.01 & ~ & 1789 & n.a. & ~ \\ 
        Student with F diagnosis & 5213 & 0.04 & ~ & 1789 & 0.04 & ~ \\ 
        Predicted school performance strata & 5213 & 0.77 & 0.23 & 1789 & 0.80 & 0.21 \\ \\[-0.5ex] \hline \hline 
    \end{tabular}
\end{table}

\begin{table}[!ht]
    \centering \caption{Descriptive Statistics for Full and Control Samples:  Part 2}\label{tab_app_desc_12} \scriptsize
    \begin{tabular}{lllllll}
    \hline \hline
        ~ & \multicolumn{3}{c}{All students} &  \multicolumn{3}{c}{Control group}  \\  
        ~ & $n$ & mean & sd & $n$ & mean & sd \\ \hline \\[-0.5ex]
        \textbf{Parental characteristics} & ~ & ~ & ~ & ~ & ~ & ~ \\ 
        Mother with F diagnosis & 5213 & 0.11 & ~ & 1789 & 0.11 & ~ \\ 
        Father with F diagnosis & 5213 & 0.10 & ~ & 1789 & 0.08 & ~ \\ 
        Mother with I diagnosis & 5213 & 0.15 & ~ & 1789 & 0.14 & ~ \\ 
        Father with I diagnosis & 5213 & 0.12 & ~ & 1789 & 0.12 & ~ \\ 
        Mother with J diagnosis & 5213 & 0.13 & ~ & 1789 & 0.15 & ~ \\ 
        Father with J diagnosis & 5213 & 0.13 & ~ & 1789 & 0.13 & ~ \\ 
        Mother's education: No college & 5213 & 0.61 & ~ & 1789 & 0.62 & ~ \\ 
         - Compulsory school & 5213 & 0.15 & ~ & 1789 & 0.15 & ~ \\ 
         - High school & 5213 & 0.06 & ~ & 1789 & 0.06 & ~ \\ 
         - Vocational & 5213 & 0.35 & ~ & 1789 & 0.35 & ~ \\ 
         - Missing & 5213 & 0.05 & ~ & 1789 & 0.05 & ~ \\ 
        Mother's education: College & 5213 & 0.34 & ~ & 1789 & 0.33 & ~ \\ 
         - Short cycle & 5213 & 0.05 & ~ & 1789 & 0.06 & ~ \\ 
         - Medium cycle & 5213 & 0.24 & ~ & 1789 & 0.23 & ~ \\ 
         - Long cycle & 5213 & 0.10 & ~ & 1789 & 0.11 & ~ \\ 
        Father's education: No college & 5213 & 0.68 & ~ & 1789 & 0.68 & ~ \\ 
         - Compulsory school & 5213 & 0.17 & ~ & 1789 & 0.16 & ~ \\ 
         - High school & 5213 & 0.05 & ~ & 1789 & 0.05 & ~ \\ 
         - Vocational & 5213 & 0.39 & ~ & 1789 & 0.38 & ~ \\ 
         - Missing & 5213 & 0.07 & ~ & 1789 & 0.08 & ~ \\ 
        Mother's education: College & 5213 & 0.24 & ~ & 1789 & 0.25 & ~ \\ 
         - Short cycle & 5213 & 0.08 & ~ & 1789 & 0.08 & ~ \\ 
         - Medium cycle & 5213 & 0.14 & ~ & 1789 & 0.14 & ~ \\ 
         - Long cycle & 5213 & 0.11 & ~ & 1789 & 0.11 & ~ \\ 
        Mother: Earnings (DKK) & 5213 & 257218 & 182228 & 1789 & 254233 & 182952 \\ 
        Father: Earnings (DKK) & 5213 & 346266 & 279667 & 1789 & 343508 & 283198 \\ 
        Mother: Social benefits (DKK) & 5213 & 5531 & 24817 & 1789 & 5693 & 25322 \\ 
        Father: Social benefits (DKK) & 5213 & 3102 & 18472 & 1789 & 3492 & 19842 \\ 
        Mother: Disposable income (DKK) & 5213 & 243321 & 125217 & 1789 & 240353 & 101392 \\ 
        Father: Disposable income (DKK) & 5213 & 269271 & 260948 & 1789 & 260533 & 176714 \\ 
        Mother: Work experience (years) & 5213 & 12.82 & 7.68 & 1789 & 12.69 & 7.78 \\ 
        Father: Work experience (years) & 5213 & 16.04 & 8.84 & 1789 & 15.97 & 8.87 \\ 
        Mother's employment status  & ~ & ~ & ~ & ~ & ~ & ~ \\ 
         - Employed & 5213 & 0.79 & ~ & 1789 & 0.78 & ~ \\ 
         - Unemployed & 5213 & 0.04 & ~ & 1789 & 0.05 & ~ \\ 
         - Inactive & 5213 & 0.15 & ~ & 1789 & 0.14 & ~ \\ 
         - Missing & 5213 & 0.02 & ~ & 1789 & 0.03 & ~ \\ 
        Father's employment status  & ~ & ~ & ~ & ~ & ~ & ~ \\ 
         - Employed & 5213 & 0.82 & ~ & 1789 & 0.81 & ~ \\ 
         - Unemployed & 5213 & 0.04 & ~ & 1789 & 0.04 & ~ \\ 
         - Inactive & 5213 & 0.09 & ~ & 1789 & 0.10 & ~ \\ 
         - Missing & 5213 & 0.05 & ~ & 1789 & 0.05 & ~ \\ 
         \\[-0.5ex] \hline \hline
    \end{tabular}
\end{table}

\begin{table}[!ht]
    \centering \caption{Descriptive Statistics for Co-teacher and Teaching Assistant Samples:  Part 1}\label{tab_app_desc_21} \scriptsize
    \begin{tabular}{lllllll}
    \hline \hline
        ~ & \multicolumn{3}{c}{Co-teacher} &  \multicolumn{3}{c}{Teaching Assistant}  \\  
        ~ & $n$ & mean & sd & $n$ & mean & sd \\ \hline \\[-0.5ex]
        \textbf{Student outcomes} & ~ & ~ & ~ & ~ & ~ & ~ \\ 
        SDQ internalizing score & 1624 & 4.38 & 3.25 & 1423 & 4.25 & 3.18 \\ 
        SDQ internalizing score, missing  & 1814 & 0.105 & ~ & 1610 & 0.116 & ~ \\ 
        SDQ externalizing score & 1624 & 5.07 & 3.50 & 1423 & 4.91 & 3.46 \\ 
        SDQ externalizing score, missing & 1814 & 0.105 & ~ & 1610 & 0.116 & ~ \\ 
        ~ & ~ & ~ & ~ & ~ & ~ & ~ \\ 
        \textbf{Subgroup indicators} & ~ & ~ & ~ & ~ & ~ & ~ \\ 
        SES education: Yes & 1814 & 0.41 & ~ & 1610 & 0.43 & ~ \\ 
        SES education: No/m & 1814 & 0.59 & ~ & 1610 & 0.57 & ~ \\ 
        SES employment: Both & 1814 & 0.70 & ~ & 1610 & 0.70 & ~ \\ 
        SES employment: One & 1814 & 0.20 & ~ & 1610 & 0.20 & ~ \\ 
        SES employment: None/m & 1814 & 0.10 & ~ & 1610 & 0.09 & ~ \\ 
        ~ & ~ & ~ & ~ & ~ & ~ & ~ \\ 
        \textbf{Student birth information} & ~ & ~ & ~ & ~ & ~ & ~ \\ 
        Boy & 1814 & 0.50 & ~ & 1610 & 0.52 & ~ \\ 
        Birthweight (grams) & 1814 & 3292 & 1048 & 1610 & 3269 & 1080 \\ 
        Birthweight low (0/1) & 1814 & 0.06 & ~ & 1610 & 0.06 & ~ \\ 
        Birthweight high  (0/1) & 1814 & 0.04 & ~ & 1610 & 0.04 & ~ \\ 
        Premature week 32 (0/1) & 1814 & 0.01 & ~ & 1610 & 0.01 & ~ \\ 
        Mother's age at birth & 1814 & 28.79 & 6.27 & 1610 & 28.57 & 6.48 \\ 
        Father's age at birth & 1814 & 30.40 & 9.01 & 1610 & 30.43 & 8.86 \\ 
        Birth order by mother & 1814 & 1.84 & 1.04 & 1610 & 1.79 & 0.99 \\ 
        Birth order by father & 1814 & 1.84 & 1.09 & 1610 & 1.79 & 1.03 \\ 
        Twin, triplet etc. & 1814 & 0.03 & ~ & 1610 & 0.04 & ~ \\ 
        Student demographics & ~ & ~ & ~ & ~ & ~ & ~ \\ 
        Origin: Danish & 1814 & 0.87 & ~ & 1610 & 0.88 & ~ \\ 
        Origin: Nonwestern & 1814 & 0.11 & ~ & 1610 & 0.11 & ~ \\ 
        Origin: Western & 1814 & 0.01 & ~ & 1610 & 0.01 & ~ \\ 
        Origin: missing & 1814 & 0.01 & ~ & 1610 & 0.00 & ~ \\ 
        Living with both parents & 1814 & 0.65 & ~ & 1610 & 0.66 & ~ \\ 
        Living with single mother & 1814 & 0.24 & ~ & 1610 & 0.23 & ~ \\ 
        Placed out of home & 1814 & 0.01 & ~ & 1610 & 0.01 & ~ \\ 
        \\
        \textbf{Student school information} & ~ & ~ & ~ & ~ & ~ & ~ \\ 
        Reading score from grade 4 & 1728 & 0.04 & 0.98 & 1549 & 0.05 & 0.97 \\ 
        Reading score, missing & 1814 & 0.05 & ~ & 1610 & 0.04 & ~ \\ 
        Math score from grade 3 & 1578 & -0.02 & 0.98 & 1403 & 0.09 & 1.03 \\ 
        Math score, missing & 1814 & 0.13 & ~ & 1610 & 0.13 & ~ \\ 
        Special education needs & 1814 & 0.13 & ~ & 1610 & 0.12 & ~ \\ 
        Special education class & 1814 & 0.01 & ~ & 1610 & 0.01 & ~ \\ 
        Student with F diagnosis & 1814 & 0.04 & ~ & 1610 & 0.05 & ~ \\ 
        Predicted school performance strata & 1814 & 0.74 & 0.25 & 1610 & 0.78 & 0.24 \\  \\[-0.5ex] \hline \hline
    \end{tabular}
\end{table}

\begin{table}[!ht]
    \centering \caption{Descriptive Statistics for Co-teacher and Teaching Assistant Samples:  Part 2}\label{tab_app_desc_22}\scriptsize
    \begin{tabular}{lllllll}
    \hline \hline
        ~ & \multicolumn{3}{c}{Co-teacher} &  \multicolumn{3}{c}{Teaching Assistant}  \\  
        ~ & $n$ & mean & sd & $n$ & mean & sd \\ \hline \\[-0.5ex] 
        \textbf{Parental characteristics} & ~ & ~ & ~ & ~ & ~ & ~ \\ 
        Mother with F diagnosis & 1814 & 0.12 & ~ & 1610 & 0.11 & ~ \\ 
        Father with F diagnosis & 1814 & 0.11 & ~ & 1610 & 0.09 & ~ \\ 
        Mother with I diagnosis & 1814 & 0.16 & ~ & 1610 & 0.14 & ~ \\ 
        Father with I diagnosis & 1814 & 0.12 & ~ & 1610 & 0.12 & ~ \\ 
        Mother with J diagnosis & 1814 & 0.12 & ~ & 1610 & 0.12 & ~ \\ 
        Father with J diagnosis & 1814 & 0.13 & ~ & 1610 & 0.14 & ~ \\ 
        Mother's education: No college & 1814 & 0.62 & ~ & 1610 & 0.61 & ~ \\ 
         - Compulsory school & 1814 & 0.17 & ~ & 1610 & 0.14 & ~ \\ 
         - High school & 1814 & 0.07 & ~ & 1610 & 0.06 & ~ \\ 
         - Vocational & 1814 & 0.33 & ~ & 1610 & 0.35 & ~ \\ 
         - Missing & 1814 & 0.04 & ~ & 1610 & 0.05 & ~ \\ 
        Mother's education: College & 1814 & 0.34 & ~ & 1610 & 0.35 & ~ \\ 
         - Short cycle & 1814 & 0.05 & ~ & 1610 & 0.06 & ~ \\ 
         - Medium cycle & 1814 & 0.24 & ~ & 1610 & 0.26 & ~ \\ 
         - Long cycle & 1814 & 0.10 & ~ & 1610 & 0.09 & ~ \\ 
        Father's education: No college & 1814 & 0.69 & ~ & 1610 & 0.68 & ~ \\ 
         - Compulsory school & 1814 & 0.18 & ~ & 1610 & 0.15 & ~ \\ 
         - High school & 1814 & 0.05 & ~ & 1610 & 0.05 & ~ \\ 
         - Vocational & 1814 & 0.38 & ~ & 1610 & 0.39 & ~ \\ 
         - Missing & 1814 & 0.07 & ~ & 1610 & 0.07 & ~ \\ 
        Mother's education: College & 1814 & 0.23 & ~ & 1610 & 0.25 & ~ \\ 
         - Short cycle & 1814 & 0.08 & ~ & 1610 & 0.09 & ~ \\ 
         - Medium cycle & 1814 & 0.13 & ~ & 1610 & 0.14 & ~ \\ 
         - Long cycle & 1814 & 0.11 & ~ & 1610 & 0.11 & ~ \\ 
        Mother: Earnings (DKK) & 1814 & 260857 & 185353 & 1610 & 256436 & 177862 \\ 
        Father: Earnings (DKK) & 1814 & 352040 & 285128 & 1610 & 342823 & 269385 \\ 
        Mother: Social benefits (DKK) & 1814 & 5636 & 25029 & 1610 & 5234 & 24008 \\ 
        Father: Social benefits (DKK) & 1814 & 3063 & 18445 & 1610 & 2713 & 16856 \\ 
        Mother: Disposable income (DKK) & 1814 & 244922 & 108952 & 1610 & 244815 & 161167 \\ 
        Father: Disposable income (DKK) & 1814 & 281438 & 364686 & 1610 & 265273 & 189071 \\ 
        Mother: Work experience (years) & 1814 & 12.84 & 7.65 & 1610 & 12.94 & 7.62 \\ 
        Father: Work experience (years) & 1814 & 15.97 & 8.86 & 1610 & 16.20 & 8.78 \\ 
        Mother's employment status  & ~ & ~ & ~ & ~ & ~ & ~ \\ 
         - Employed & 1814 & 0.79 & ~ & 1610 & 0.79 & ~ \\ 
         - Unemployed & 1814 & 0.04 & ~ & 1610 & 0.03 & ~ \\ 
         - Inactive & 1814 & 0.15 & ~ & 1610 & 0.15 & ~ \\ 
         - Missing & 1814 & 0.02 & ~ & 1610 & 0.02 & ~ \\ 
        Father's employment status  & ~ & ~ & ~ & ~ & ~ & ~ \\ 
         - Employed & 1814 & 0.82 & ~ & 1610 & 0.82 & ~ \\ 
         - Unemployed & 1814 & 0.04 & ~ & 1610 & 0.04 & ~ \\ 
         - Inactive & 1814 & 0.09 & ~ & 1610 & 0.09 & ~ \\ 
         - Missing & 1814 & 0.05 & ~ & 1610 & 0.05 & ~ \\ 
          \\[-0.5ex] \hline \hline
    \end{tabular}
\end{table}

\end{document}